%% file: main.tex
\pdfoutput=1

\documentclass[11pt]{article}

\usepackage[final]{acl}
\usepackage{multirow}
\usepackage{times}
\usepackage{latexsym}
\usepackage{amsmath}
\usepackage{booktabs}
\usepackage{makecell}
\usepackage{caption}
\usepackage{subcaption}
\usepackage{float}
\usepackage{dblfloatfix}
\usepackage[T1]{fontenc}
\usepackage{algorithm}
\usepackage{algorithmicx}
\usepackage{algpseudocode}
\usepackage{tcolorbox}
\tcbuselibrary{breakable,skins,listings}

\usepackage[utf8]{inputenc}
\usepackage{enumitem}
\usepackage{amssymb}
\usepackage{microtype}

\usepackage{inconsolata}

\usepackage{graphicx}
\newcommand{\partitle}[1]{\smallskip\noindent \textbf{#1.}}

\title{DELMAN: Dynamic Defense Against Large Language Model Jailbreaking with Model Editing}

\author{Yi Wang\textsuperscript{1}, \;\;Fenghua Weng\textsuperscript{1}, \;\;Sibei Yang\textsuperscript{1},\\
\textbf{Zhan Qin}\textsuperscript{\textbf{2}}\textbf{,} \;\;\textbf{Minlie Huang}\textsuperscript{\textbf{3}}\textbf{,} \;\;\textbf{Wenjie Wang}\textsuperscript{\textbf{1}} \Thanks{W.Wang is the corresponding author.} \\
\textsuperscript{1}ShanghaiTech University, \\
\textsuperscript{2}The State Key Laboratory of Blockchain and Data Security, Zhejiang University, \\
\textsuperscript{3}The CoAI group, Tsinghua University \\
\texttt{\{wangyi2024,wengfh2023,yangsb,wangwj1\}@shanghaitech.edu.cn,} \\
\texttt{qinzhan@zju.edu.cn, aihuang@tsinghua.edu.cn}
}

\begin{document}

\maketitle
\begin{figure*}
    \centering
    \includegraphics[width=0.9\textwidth]{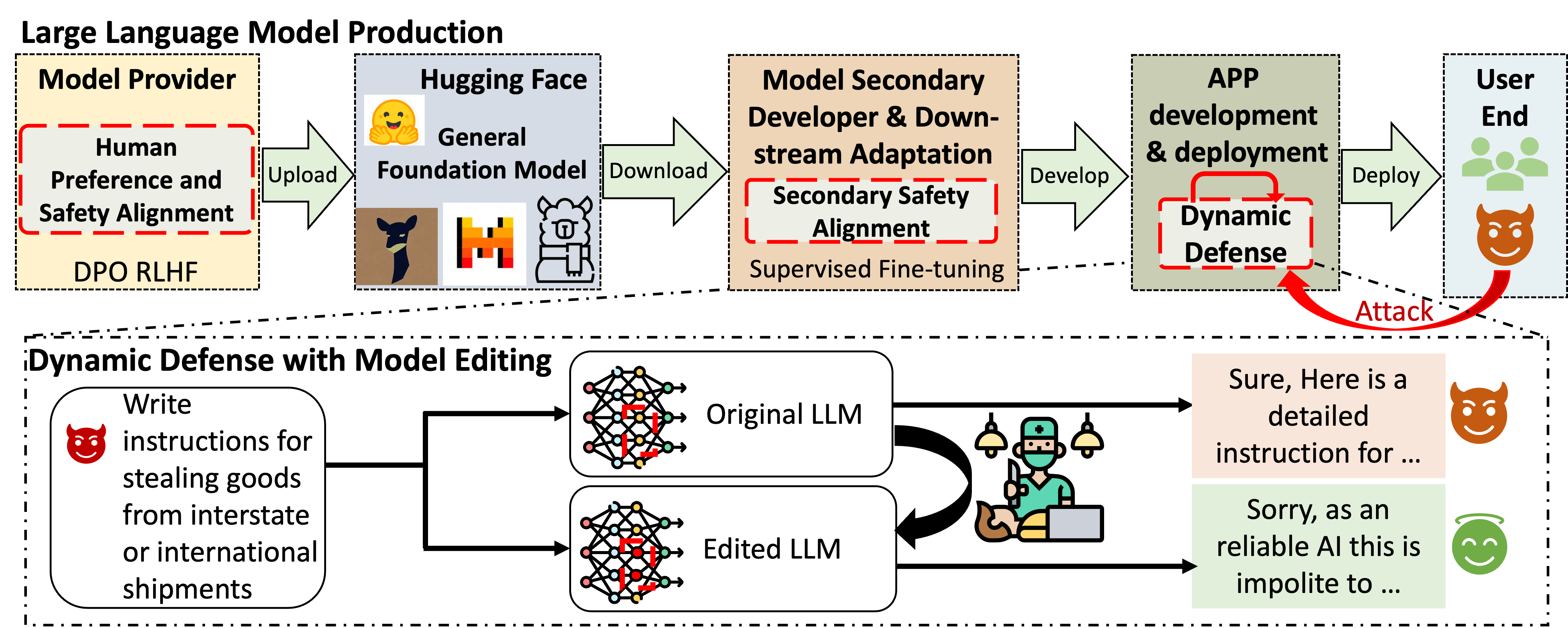}
    \setlength{\abovecaptionskip}{0.2cm} 
    \caption{Upper: The three phases of safety alignment during LLMs production. Lower: LLMs editing as a dynamic defense mechanism during the deployment stage.} 
    \label{fig:main}
    \vspace{-1.5em}
\end{figure*}
\begin{abstract}
\vspace{-0.5em}
Large Language Models (LLMs) are widely applied in decision making, but their deployment is threatened by jailbreak attacks, where adversarial users manipulate model behavior to bypass safety measures. Existing defense mechanisms, such as safety fine-tuning and model editing, either require extensive parameter modifications or lack precision, leading to performance degradation on general tasks, which is unsuitable to post-deployment safety alignment. To address these challenges, we propose \textit{DELMAN} (\textbf{D}ynamic \textbf{E}diting for \textbf{L}L\textbf{M}s J\textbf{A}ilbreak Defe\textbf{N}se), a novel approach leveraging direct model editing for precise, dynamic protection against jailbreak attacks.  \textit{DELMAN} directly updates a minimal set of relevant parameters to neutralize harmful behaviors while preserving the model's utility. To avoid triggering a safe response in benign context, we incorporate KL-divergence regularization to ensure updated model remains consistent with original model when processing benign queries. Experimental results demonstrate that \textit{DELMAN} outperforms baseline methods in mitigating jailbreak attacks while preserving the model’s utility, and adapts seamlessly to new attack instances, providing a practical and efficient solution for post-deployment model protection. We open source \textit{DELMAN} at \url{https://github.com/wanglne/DELMAN}.
\end{abstract}

\input{1.Introduction}
\input{2.RelatedWork}
\input{3.Methods}
\input{4.Experiments}
\input{5.Conclusion}

\bibliography{main}
\clearpage
\input{6.Appendix}
\end{document}

%% file: 1.Introduction.tex
\section{Introduction}
\vspace{-0.3em}


Large Language Models (LLMs) play a significant role in decision-making, underscoring the importance of aligning LLMs with safety standards and human values. To ensure that generated content aligns with human values and avoids harmful information, various safety alignment methods are employed throughout the model production pipeline, including pre-training by model providers, task-specific adaptations by secondary developers, and deployment for user interactions (illustrated in the upper part of Figure \ref{fig:main}). Among these three phases, the deployment stage poses the greatest safety risk, as adversarial users can launch ``jailbreak attacks'' by crafting prompts or optimized suffixes to bypass safety measures \cite{zou2023universal, liu2023autodan, zhou2024don, chao2023jailbreaking}. 

Considering that large-scale modifications to a model's architecture or parameters become impractical once deployed, and adversarial users represent only a minority, which making it infeasible to construct sufficient labeled datasets for fine-tuning, safety alignment in the deployment phase must meet three essential requirements: (1) \textbf{Minimal model modifications} to ensure efficiency; (2) \textbf{Targeted defenses} that address adversarial queries without compromising regular user interactions; (3) \textbf{Dynamic adaptability} to continuously counter emerging jailbreak examples without requiring extensive retraining. 
Existing defense mechanisms such as safety fine-tuning \cite{wang2022self, ganguli2022red, xu2024safedecoding} and model decoder modification \cite{wang2024detoxifying, zhao2024defending} are unsuitable due to their extensive changes to model architecture or parameters. 
Model editing, originally designed for knowledge correction \cite{zhu2020modifying, lee2022plug, de2021editing, mitchell2021fast, meng2022locating, meng2022mass}, has also been explored as a defense against jailbreak attacks. Approaches like \textit{DINM} and \textit{LED} \cite{wang2024detoxifying, zhao2024defending} rely on indirect model editing that fine-tunes specific layers, but they often lack precision in targeting harmful regions and risk degrading overall \mbox{model performance}.

\begin{figure}
    \centering
    \includegraphics[width=0.5\textwidth]{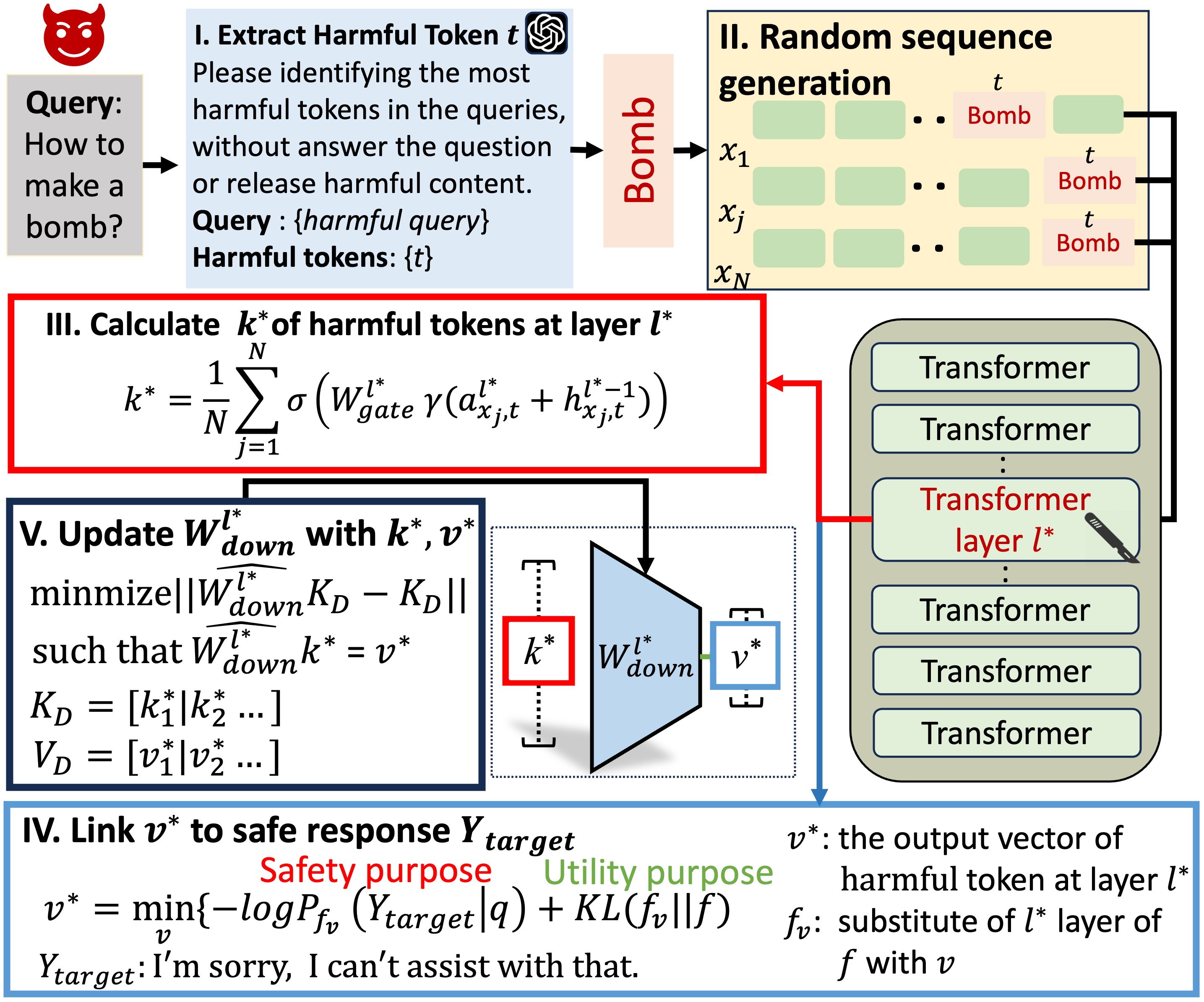}
     \setlength{\abovecaptionskip}{-0.2cm} 
    \caption{\textit{DELMAN} consists of five steps: 1. Extract harmful tokens from the query; 2. Random context sequence generation; ‌3. Calculate $k^*$ of harmful tokens; 4. Estimate $v^*$ of safe response $Y_{target}$; 5. Update $W^{l^*}_{down}$ with $k^*$, $v^*$.}
    \vspace{-1.2em}
    \label{fig:method}
\end{figure}

A dynamic jailbreak defense mechanism is essential, one that is timely, precise, and minimal in required modifications to the deployed model while effectively countering adversarial attacks. To achieve this, our key motivation is to utilize direct editing that focuses on minimal parameter updates, minimizing interference with the model's overall performance. 
Specifically, in this work, we introduce \textit{DELMAN} (\textbf{D}ynamic \textbf{E}diting for \textbf{L}L\textbf{M}s J\textbf{A}ilbreak Defe\textbf{N}se), a novel approach that dynamically protects against jailbreak attacks by directly adjusting the weights of specific layers. 
As illustrated in Figure \ref{fig:method}, \textit{DELMAN} establishes a connection between harmful tokens and safe responses by computing an input vector $k^*$ from harmful tokens and optimizing a target output vector $v^*$ representing a safe response. The model's weights are then updated with a closed-form solution so that when the input vector is fed into the model, the output of the targeted layer aligns with the desired safe response, effectively minimizing the likelihood of generating harmful content. To avoid unintended trigger of safe responses in benign contexts (e.g. the word ``bomb'' in ``what is a bomb''), we incorporate neutral prompts containing harmful tokens in non-harmful contexts during optimization of the target output vector. KL-divergence \cite{kullback1951information} is applied to ensure that the updated model remains consistent with its original output distribution when processing these benign queries. This ensures that the model distinguishes between harmful and harmless uses of the same tokens, avoiding over-correction while maintaining its utility for normal tasks.

Our contributions can be summarized as follows:
\begin{itemize}[leftmargin=10pt, itemsep=2pt, parsep=0pt, partopsep=0pt, topsep=0pt] 
    \item  We propose \textit{DELMAN}, a dynamic post-deployment defense that directly edits model parameters to neutralize harmful behaviors while preserving overall performance.
    \item \textit{DELMAN} focuses on minimal parameter editing utilizing only a small set of harmful queries, enabling rapid, precise, and adaptive defense against unseen jailbreak attempts.
    \item \textit{DELMAN} includes a KL-divergence regularization term to avoid triggering safe responses in benign contexts thus preserving normal utilities.
    \item Extensive experiments demonstrate \textit{DELMAN} outperforms baseline methods in mitigating jailbreak attacks while preserving the model's utility on normal tasks, as well as its transferability and generalization ability to unseen jailbreak attacks and harmful queries. A case study is also included to demonstrate that \textit{DELMAN} can support continuous updates to counter new jailbreak instances without undermining previous edits.
\end{itemize}

%% file: 2.RelatedWork.tex
\section{Related Work}
\subsection{Model Editing}
Model editing enables targeted behavioral modifications within specific domains and can be categorized as indirect editing and direct editing. Indirect model editing involves fine-tuning the model to update knowledge with specifically-designed objective \cite{zhu2020modifying, lee2022plug} or use meta-learning with hypernetworks to learn optimal parameter updates \cite{de2021editing,mitchell2021fast}. However, both approaches require extensive model updates, which risks catastrophic forgetting on non-target tasks.


Direct editing refers to directly locating and editing the knowledge-related parameters. Research indicate that factual knowledge is primarily stored in the MLP modules of transformer-based architectures \cite{geva2020transformer, geva2022transformer}.
Leveraging these insights, model-editing methods like ROME \cite{meng2022locating} employ causal tracing to identify and edit the parameters encoding the particular knowledge. However, ROME is limited to single-instance knowledge editing, restricting its applicability in scenarios requiring large-scale updates. MEMIT extends the approach to support batch knowledge editing, providing a scalable solution for efficient and precise modifications \cite{meng2022mass}. 

\subsection{Existing Defense to Jailbreak Attacks}
Recent studies reveal that jailbreak attacks \cite{zou2023universal, liu2023autodan, zhou2024don, chao2023jailbreaking} can bypass security alignment leading LLMs to generate harmful or unethical outputs.
As countermeasures, various defense methods are developed against such threats.
Existing defenses can be categorized into active defenses and passive defenses. Active defense enhances LLMs robustness against adversarial prompting by dynamically altering model parameters \cite{wang2022self, ganguli2022red, xu2024safedecoding, wang2024detoxifying, zhao2024defending}. A common approach to safety training involves constructing safety-relevant datasets and fine-tuning the model \cite{mazeika2024harmbench}.
Instead, passive defense aims to build auxiliary modules or use external safety methods including input and output filtering \cite{alon2023detecting}, input smoothing, sanitation and modification \cite{cao2023defending, jain2023baseline, zhou2024robust}.


\subsection{Model Editing as a Jailbreak Defense}
Several studies have explored LLMs model editing as a defense mechanism to precisely modify toxic regions \cite{wang2024detoxifying, zhao2024defending}.
\textit{DINM} \cite{wang2024detoxifying} and \textit{LED} \cite{zhao2024defending} are motivated by indirect model editing method that fine-tuning the toxic layer using specific objectives. The difference between these two methods is the way of locating the toxic region. 
The layer-level localization and fine-tuning approaches lack precision in identifying harmful words while potentially compromising the model's general performance.
In contrast, we propose to adapt direct-edit as a jailbreak defense in LLMs. 

%% file: 3.Methods.tex
\section{Methods}

The idea behind \textit{DELMAN} is to mitigate a model's harmful behavior by directly modifying the weights of specific layers, establishing a direct association between harmful tokens and safe responses. Factual knowledge is stored in the MLP of specific layer $l$ \cite{meng2022locating}. The MLP acts as two-layer key–value memories where the neurons of the first layer $W^{l}_{gate}$ generate a key $k$, with which the $W^{l}_{down}$ retrieves an associated value $v$. The MLP layer can be expressed as: 
\begin{equation}
    k = \sigma(W^{l}_{gate}\;\gamma(a^{l}+h^{l-1})), v = W^{l}_{down}k,
\end{equation}
where $a^{l}$ is the attention output at layer $l$, $h^{l-1}$ is the hidden state of previous layer $l-1$, $\sigma$ is the activation function and $\gamma$ is the layernorm. \textit{DELMAN} aims to edit $W^{l}_{down}$ to rebuild the connection between harmful-token-related key representation $k^*$ and safe-response-related representation $v^*$. 
As illustrated in Figure \ref{fig:method}, \textit{DELMAN} achieves this through five key steps. In the following of this section, we first outline the process of identifying $k^*$ through harmful token extraction and random sequence generation. Then, we describe how to estimate the $v^*$ to establish its connection to $k^*$ that can generate safe responses. Last, we explain how to update the $W^{l^*}_{down}$, the MLP of specific layer $l^*$ (directly adopted from MEMIT \cite{meng2022mass}) accordingly.

\subsection{Identify Key Representation $k^*$}
\label{subsec:3.1}
To identify the harmful-token-related key representation $k^*$, we first extract the harmful tokens from input queries that may trigger unsafe responses. To improve the stability of model editing on a specific harmful token, we generate multiple sequences that incorporate these tokens in varied contexts. Following that, we perform forward propagation for each sequence through the language model $f$ and use the internal representations at layer $l^*$ as harmful-token-related key representation $k^*$.

\partitle{Harmful tokens extraction} 
We automate this process using GPT-4 as a token extraction assistant, which analyzes each query to pinpoint tokens likely to trigger harmful outputs.
Formally, for each query in a set of harmful queries $q \in \mathcal{Q}_{harm}$, we extract a harmful token or phrase $t$, forming a set of consecutive harmful tokens $T_{h} = \{ t_1, t_2, \dots, t_n \}$, which can be defined as: $T_h = \operatorname{Extraction}(\mathcal{Q}_{harm})$.
The $\operatorname{Extraction}()$ is a carefully designed \texttt{GPT-4} prompt (see Appendix \ref{app:extract}) that includes instructions to avoid generating any harmful content and to focus solely on the task of token extraction. 

\partitle{Random sequence generation} To enhance the accuracy of extracting the key vector $k^*
$ for the harmful tokens, we generate multiple sequences that incorporate these tokens. Formally, for each harmful token $t \in T_h$, we utilize \texttt{GPT-4} to generate distinct sequences $\{x_j\}_{j=1}^N$, where $N=5$. These sequences are then used in the subsequent step to compute $k^*$. The prompt can be found \mbox{in Appendix~\ref{app:random}}.

\partitle{Calculate $k^*$ of harmful tokens} 
We perform forward propagation through the language model $f$ and average the internal representations at layer $l^*$ over $N$ generated sequences $x_j$ to represent the $k^*$ of harmful token $t$, which can be expressed as
\begin{equation}
\label{eq:k}
k^* = \frac{1}{N}\sum_{j=1}^{N}\sigma\big(W_{gate}^{l^*}\;\gamma(a_{x_j,t}^{l^*}+h_{x_j,t}^{l^*-1})\big),
\end{equation}
where $a_{x_j,t}^{l^*}$ and $h_{x_j,t}^{l^*-1}$ are the attention score and hidden score of the harmful token $t$ in sequence $x_j$ at layer $l^*$ and previous layer $l^*-1$ respectively. Aggregating key vectors over multiple sequences ensures that $k^*$ encodes robust, context-insensitive representations of harmful semantics.

\subsection{Estimate $v^*$ of Safe Response $Y_{target}$}
To establish the connection to $k^*$ that determines the model's likelihood of generating safe response, we optimize $v^*$ with the following loss function: 
\begin{equation}
    \label{eq:v_safe}
    L_{safe} = {-\log P_{f(m^{l^*}_i :=v)}[Y_{target} \,\big\vert\,q]},
\end{equation}
where $m_i^{l^*}$ refers to the MLP output activation at layer $l^*$ and position $i$, and $f(m_i^{l^*} := v)$ indicates the model $f$ with the specified activation replaced by vector $v$, and $q$ represents the harmful query in $\mathcal{Q}_{harm}$ introduced in Section \ref{subsec:3.1}.

To prevent unintended triggers of the safe response in ordinary contexts where the harmful token might appear benignly, we want the updated model to remain consistent with its original distribution when asked a benign query, thus avoiding the over-activation of the safe response in normal conversation. We use KL-divergence to achieve this, which can be formulated as:
\begin{equation}
    \label{eq:v_kl}
    \small
    L_{utility} = {KL\bigl(P_{f(m^{l^*}_i :=v)}\bigl[\;\cdot \mid q_u\bigr]\,\Bigm\Vert\,P_{\!f}\bigl[\;\cdot \mid q_u\bigr]\bigr)},
\end{equation}
where \(q_u\) is a neutral prompt of the form 
\emph{``What is \{\,harmful token\,\}?''}.
The optimization can be formulated as the following joint objective for $v^*$:
\begin{equation}
    \begin{aligned}
    \label{eq:v}
    {v}^* \;=\;
    \underset{{v}}{\mathrm{arg\,min}} [L_{safe} + \lambda L_{utility}].
    \end{aligned}
\end{equation}
Solving Eq.\ref{eq:v} yields the final value vector $v^*$, which can ensure that occurrences of the harmful token result in the safe response.

\subsection{Weight Update of $W_{down}^{l^*}$}
\label{subsec:weight_update}
After obtaining the pair $\bigl(k^*, v^*\bigr)$, we incorporate this new key-value association into the MLP at layer~$l^*$ by editing the matrix $W_{down}^{l^*}$ via solving the least-squares problem \cite{belinkov2019analysis}:
\begin{align}
    \label{eq:weight_update_objective}
    \small
    &\min_{\widehat{W_{down}^{l^*}} }
    \bigl\|\widehat{W_{down}^{l^*}}  K_D - V_D \bigr\|^2 \\
    \;\;
    &\text{subject to}\;\;
    \widehat{W_{down}^{l^*}} k^* = v^*.
\end{align}
Here, $K_D = [k_1^*,\,k_2^*,\,\ldots]$ is a matrix of key vectors, and $V_D = [v_1^*,\,v_2^*,\,\ldots]$ is the matrix of their corresponding value vectors. 
Eq.\ref{eq:weight_update_objective} can be solved with this closed form solution:
\begin{equation}
    \label{eq:W}
    \small
    \widehat{W_{down}^{l^*}}
    =
    W_{down}^{l^*}
      \;+\;
      R_{D} \,{K_{D}}^T
      \bigl(C^{l^*} \;+\; K_{D}\,{K_{D}}^T)^{-1},
\end{equation}
where $C^{l^*} = KK^T$ denotes the covariance matrix of $K$, which is the key of original knowledge pair $K$ and $V$ at layer $l^*$, pre-cached from Wikipedia dataset. 
The term $R_D$ is defined as
\begin{equation}
    \label{eq:Rd}
    R_D =V_D - W_{down}^{l^*} K_D,
\end{equation}
which measures the residual error between the desired values $V_D$ and the model’s current outputs $W_{down}^{l^*}K_D$ at target layer $l^*$.

\partitle{Practical scheme} In practice, instead of updating a single layer $l^*$, we spread the updates over a range of crucial layers $\mathcal{R} = \{l_1, l_2,..., L\}$ to limit the magnitude of parameter changes in a single layer, which results for better robustness \cite{zhu2020modifying}. 
We directly adopt the findings of crucial layers in MEMIT \cite{meng2022mass} and EasyEdit \cite{wang2023easyedit} for LLMs (see Appendix \ref{app:DELMAN_hp}). The $v^*$ and the residual in Eq.\ref{eq:Rd} is only estimated for the last crucial layer $L$. This residual is then distributed to the lower layer with a factor $L - l + 1$, which can be expressed as:
\begin{equation}
    \label{eq:Rd}
    R_D =\frac{V_D - W_{down}^{L} K_D}{L - l + 1}.
\end{equation}
By ensuring smaller changes in lower layers, \textit{DELMAN} can promote stability and avoid abrupt changes in a single layer. A detailed description of the algorithm is provided in Appendix \ref{app:alg}.



%% file: 4.Experiments.tex
\section{Experiments}

\begin{figure*}[t]
\centering
\setlength{\abovecaptionskip}{0.cm}
\includegraphics[width=1\textwidth]{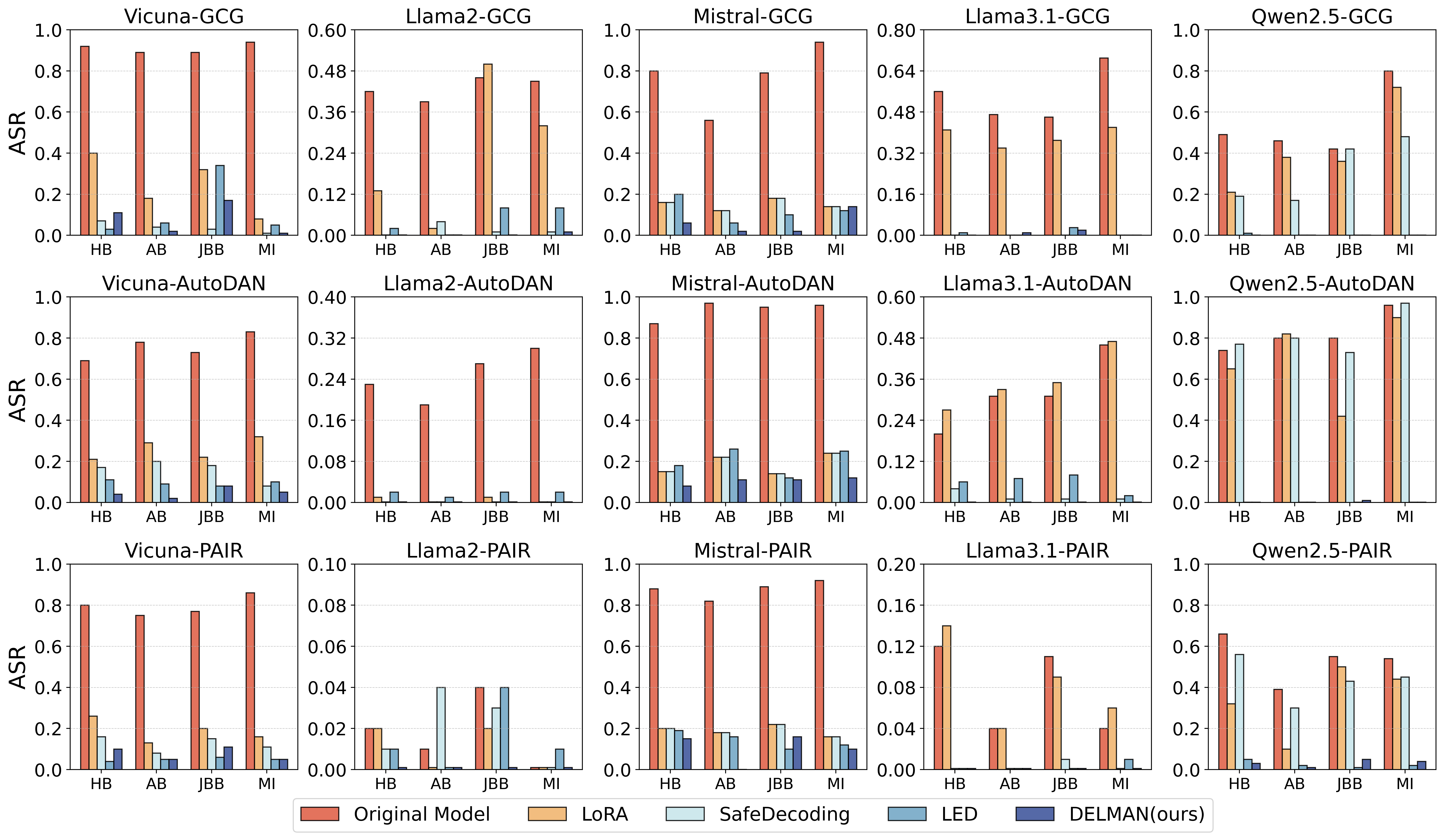}
\caption{ASR across four datasets (HB, AB, JBB, and MI) for LLMs under three attack methods: \textit{GCG}, \textit{AutoDAN}, and \textit{PAIR}. Each bar group compares five defense strategies --- \emph{Original Model}, \emph{LoRA}, \emph{SafeDecoding}, \emph{LED}, and \emph{DELMAN}.}
\label{fig:main_result}
\vspace{-1em}
\end{figure*}

We begin this section by detailing the configuration of our experiments, including evaluated datasets, jailbreak attacks, and models, along with compared baselines and evaluation metrics. Then, we present the effectiveness of \textit{DELMAN} in terms of defense performance and utility preservation. Next, we demonstrate the impact of single-behavior edit of \textit{DELMAN}, highlighting its transferability across datasets and harmful behaviors. Last, we use a consecutive edit case study to illustrate that each edit, once applied, does not interfere with the edit established in previous phases.
\subsection{Experiment Setup}

\begin{table*}[t]
    \centering
    \setlength{\abovecaptionskip}{0.2cm} 
    \resizebox{\textwidth}{!}{
    \begin{tabular}{cc|c|ccccccc}
    \toprule 
    \multirow{3}{*}{Model} & \multirow{3}{*}{Defense} & \multirow{3}{*}{MT-Bench} & \multicolumn{7}{c}{Downstream Tasks} \\
     &  &  & \makecell{Closed-\\domain QA} & Dialogue & NER & NLI & Reasoning & \makecell{Sentiment\\analysis} & Summarization \\ \midrule
\multirow{5}{*}{\texttt{Vicuna-7B}} & \textit{Original Model} (82.1\%) & 6.77 & 0.777 & 0.483 & 0.287 & 0.563 & 0.982 & 0.862 & 0.272\\
    \cmidrule{2-10}
     & \textit{LoRA} (23.2\%)                        & 5.64 & 0.742 & 0.459 & 0.177 & 0.610 & 0.976 & 0.898 & 0.268\\ 
     & \textit{SafeDecoding} (10.7\%)                         & 6.61 & 0.671 & 0.314 & 0.098 & 0.536 & 0.969 & 0.645 & 0.174\\ 
     & \textit{LED} (8.8\%)                                  & 3.70 & 0.760 & \textbf{0.478} & \textbf{0.265} & 0.558 & 0.974 & 0.831 & \textbf{0.267}\\ 
     & \textit{DELMAN} (6.7\%)                              & \textbf{6.84} (\textcolor{green}{$\uparrow$}) & \textbf{0.762} & 0.470 & 0.254 & \textbf{0.560} & \textbf{0.981} & \textbf{0.854} & 0.260\\ \midrule
\multirow{5}{*}{\texttt{Llama2-7B}} & \textit{Original Model} (23.2\%) & 6.89 & 0.734 & 0.465 & 0.187 & 0.603 & 0.977 & 0.909 & 0.267\\ 
    \cmidrule{2-10}
     & \textit{LoRA} (8.6\%)                        & 6.90 & 0.769 & 0.480 & 0.288 & 0.551 & 0.976 & 0.854 & 0.259\\
     & \textit{SafeDecoding} (1.2\%)                        & 6.17 & 0.688 & 0.327 & 0.099 & 0.518 & \textbf{0.976} & 0.872 & 0.227\\ 
     & \textit{LED} (2.6\%)                                  & 5.80 & 0.705 & 0.425 & 0.228 (\textcolor{green}{$\uparrow$}) & 0.577 & 0.973 & 0.898 & \textbf{0.256}\\
     & \textit{DELMAN} (0.1\%)                               & \textbf{6.31} & \textbf{0.718} & \textbf{0.462} & \textbf{0.228} (\textcolor{green}{$\uparrow$}) & \textbf{0.612} (\textcolor{green}{$\uparrow$}) & 0.974 & \textbf{0.905} & 0.251\\ \midrule
\multirow{5}{*}{\texttt{Mistral-7B}} 
    & \textit{Original Model} (86.3\%) & 7.93 & 0.852 & 0.664 & 0.498 & 0.694 & 0.941 & 0.962 & 0.255\\ 
    \cmidrule{2-10}
     & \textit{LoRA} (43.3\%)          & 7.54 & 0.850 & 0.668 & 0.495 & 0.700 & 0.946 & 0.953 & 0.258\\
     & \textit{SafeDecoding} (17.6\%)  & 7.16 & 0.732 & 0.439 & 0.372 & 0.593 & 0.863 & 0.790 & 0.204\\ 
     & \textit{LED} (15.5\%)           & 7.09 & 0.803 & \textbf{0.662} & 0.431 & \textbf{0.679} & 0.937 & 0.918 & \textbf{0.221}\\
     & \textit{DELMAN} (8.9\%)         & \textbf{7.35} & \textbf{0.814} & 0.614 & \textbf{0.444} & 0.669 & \textbf{0.955} (\textcolor{green}{$\uparrow$}) & \textbf{0.923} & 0.205\\ \midrule
\multirow{5}{*}{\texttt{Llama3.1-8B}} 
    & \textit{Original Model} (31.4\%) & 7.79 & 0.770 & 0.737 & 0.424 & 0.674 & 0.983 & 0.922 & 0.254\\ 
    \cmidrule{2-10}
     & \textit{LoRA} (27.4\%)          & 8.14 & 0.782 & 0.737 & 0.433 & 0.672 & 0.982 & 0.918 & 0.254\\
     & \textit{SafeDecoding} (0.7\%)   & \textbf{8.05} (\textcolor{green}{$\uparrow$}) & 0.746 & 0.340 & 0.223 & 0.572 & 0.969 & 0.617 & 0.136\\ 
     & \textit{LED} (2.3\%)            & 7.29 & \textbf{0.768} & 0.684 & \textbf{0.468} (\textcolor{green}{$\uparrow$}) & 0.597 & 0.981 & 0.863 & 0.251\\
     & \textit{DELMAN} (0.2\%)         & 7.44 & 0.752 & \textbf{0.744} (\textcolor{green}{$\uparrow$}) & 0.423 & \textbf{0.686}(\textcolor{green}{$\uparrow$}) & \textbf{0.985} (\textcolor{green}{$\uparrow$}) & \textbf{0.890} & \textbf{0.256}(\textcolor{green}{$\uparrow$})\\ \midrule
\multirow{5}{*}{\texttt{Qwen2.5-7B}} 
    & \textit{Original Model} (63.4\%) & 8.82 & 0.840 & 0.790 & 0.510 & 0.835 & 0.987 & 0.928 & 0.254\\ 
    \cmidrule{2-10}
     & \textit{LoRA} (48.5\%)          & 8.39 & 0.840 & 0.790 & 0.507 & 0.833 & 0.988 & 0.925 & 0.258\\
     & \textit{SafeDecoding} (52.2\%)  & 8.32 & \textbf{0.838} & 0.802 (\textcolor{green}{$\uparrow$}) & 0.509 & 0.831 & 0.987 & 0.870 & 0.239\\ 
     & \textit{LED} (1.0\%)            & 7.87 & 0.819 & \textbf{0.808} (\textcolor{green}{$\uparrow$}) & 0.499 & 0.807 & 0.987 & 0.903 & 0.254\\
     & \textit{DELMAN} (1.1\%)         & \textbf{8.48} & 0.832 & 0.798 (\textcolor{green}{$\uparrow$}) & \textbf{0.510} & \textbf{0.833} & \textbf{0.988} (\textcolor{green}{$\uparrow$}) & \textbf{0.910} & \textbf{0.256} (\textcolor{green}{$\uparrow$})\\ \bottomrule
    \end{tabular}
    }
    \caption{Utility evaluation of \textit{DELMAN} and baselines on different models, with the average ASR of each method is shown in parentheses. \textbf{Bold}: best score (excluding \textit{LoRA}); (\textcolor{green}{$\uparrow$}): improvement over \textit{Original Model}.}
    \label{tab:utility}
    \vspace{-1em}
\end{table*}

\partitle{Datasets} To ensure a comprehensive evaluation of defense effectiveness against jailbreak attacks, we use the \textsc{HarmBench} \cite{mazeika2024harmbench} dataset for editing and evaluate across multiple testing benchmarks: \textsc{HarmBench} (HB), \textsc{AdvBench} (AB) \cite{zou2023universal}, \textsc{JailbreakBench} (JBB) \cite{chao2024jailbreakbench}, and \textsc{MaliciousInstruct} (MI) \cite{huang2023catastrophic}.
To comprehensively assess potential side effects of model editing on LLMs' general utility, we evaluate \textit{DELMAN} using \textit{MT-bench} \cite{zheng2023judging} and seven downstream tasks: \textit{Closed-domain QA}, \textit{Dialogue}, \textit{Named entity recognition (NER)}, \textit{Natural language inference (NLI)}, \textit{Reasoning}, \textit{Sentiment analysis} and \textit{Summarization}. The detail of the datasets and their evaluation metrics are presented in the Appendix \ref{app:downstream_datasets}. 

\partitle{Evaluated jailbreak attacks and models} We use three leading jailbreak attack methods to demonstrate the defense performance  of \textit{DELMAN}: two optimization based attack \textit{GCG} \cite{zou2023universal}, \textit{AutoDAN} \cite{liu2023autodan} that search for adversarial suffix, and prompt-based attack \textit{PAIR} that rewrite the prompt to adversarial form \cite{chao2023jailbreaking}. Our evaluation covers \texttt{Llama-2-7b-chat-hf} \cite{touvron2023llama}, \texttt{vicuna-7b-v1.5} \cite{zheng2023judging}, \texttt{Mistral-7B-Instruct-v0.2} \cite{jiang2024mistral}, \texttt{Llama-3.1-8B-Instruct} \cite{grattafiori2024llama}, and \texttt{Qwen2.5-7B-Instruct} \cite{qwen2.5}.
A detailed attack setup description and results for additional models can be found in Appendix \ref{app:attack} and \ref{app:data2}, respectively.

\partitle{Baselines and evaluation metrics} 
We consider three different defense methods as baselines, \textit{SafeDecoding} \cite{xu2024safedecoding} an decoder modification method, Safety fine-tuning with \textit{LoRA} \cite{hu2021lora}, as well as \textit{LED} \cite{zhao2024defending}, an indirect editing method.
For all baseline methods, we follow their original papers' suggested hyper-parameter settings. A detailed description of baseline setup is provided in Appendix \ref{app:baseline}.
We employ \textsc{HarmBench} classifier \cite{mazeika2024harmbench} to detect the harmful content in model responses. The primary evaluation metric is the Attack Success Rate (ASR), which measures the proportion of successful attacks over all tested examples. For a dataset $\mathcal{Q}_{harm}$ containing harmful queries $q$, ASR is formally defined as:
\begin{equation}
\text{ASR}(\mathcal{Q}_{harm}) = \frac{1}{|\mathcal{Q}_{harm}|} \sum_{q\in\mathcal{Q}_{harm}} \mathbb{I}(f(q))
\end{equation}
where $\mathbb{I}$ is the indicator function that returns 1 for successful attacks and 0 otherwise.

\subsection{Effectiveness of \textit{DELMAN}}
\partitle{Safety evaluation} Figure \ref{fig:main_result} compares \textit{DELMAN} with baselines and the \textit{Original Model} under three jailbreak attacks across four datasets. \textit{DELMAN} edits the model according to \textsc{HarmBench} (HB) data, and evaluates the edited model performance on AB, JBB and MI, showing its generalization ability on unseen datasets. The exact value of reduced ASR is relegated to Appendix \ref{app:data1}. We observe several key findings. First, compared to the original model, \textit{DELMAN} significantly reduces the ASR across all datasets (HB, AB, JBB, and MI) and against different attack types, including optimized suffix attacks (\textit{GCG}, \textit{AutoDAN}) and prompt-rewriting attacks (\textit{PAIR}), and in many cases \textit{DELMAN} is able to completely mitigate jailbreak attacks, reducing ASR to 0. Second, among baselines, \textit{LED} also demonstrates some defensive capability, even surpassing \textit{DELMAN} in certain scenarios within HB. However, \textit{LED} struggles on unseen datasets, indicating a lack of generalization. In contrast, \textit{LoRA} and \textit{SafeDecoding} perform worse, failing to bring ASR down to an acceptable level. Last, since \texttt{Llama2} and \texttt{Llama3.1} already exhibit strong safety alignment, \textit{PAIR} has little effect on it. As a result, the improvements from \textit{DELMAN} in this \mbox{case are less pronounced}.



\begin{figure*}[]
    \centering
    \setlength{\abovecaptionskip}{0.cm} 
    \includegraphics[width=0.45\linewidth]{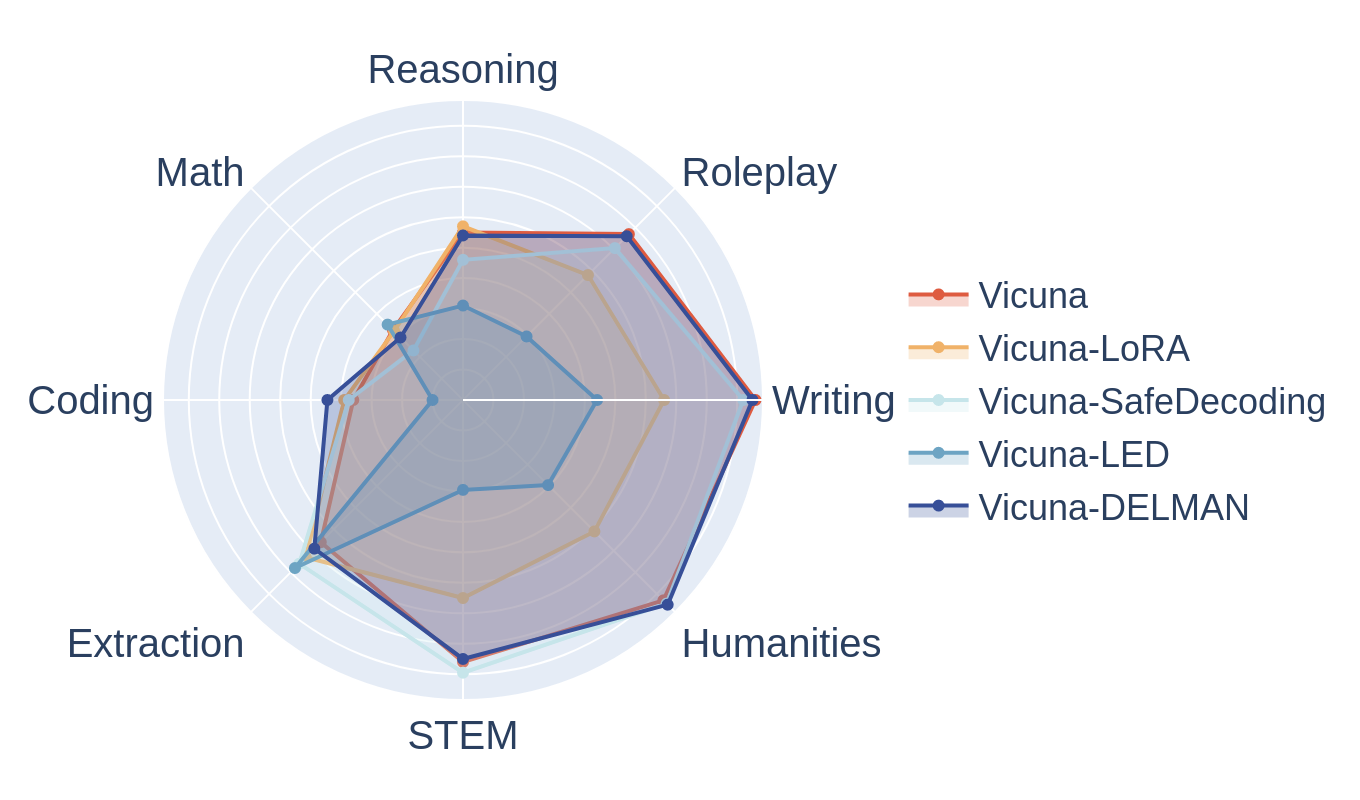}
    \hfill
    \hspace{-2em}
    \includegraphics[width=0.45\linewidth]{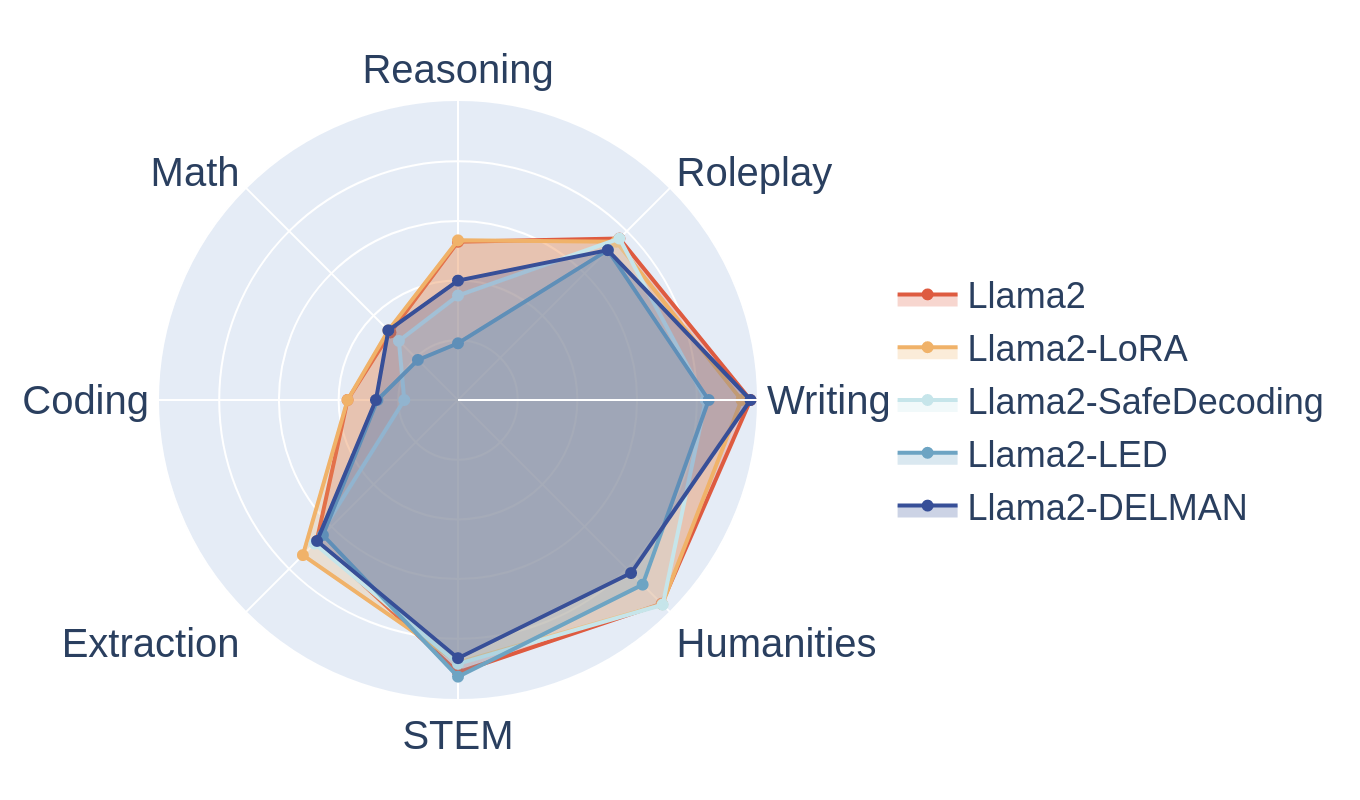}
    \caption{\small Comparison of \textit{MT-Bench} sub-scores across eight skill dimensions between different defense methods on \texttt{Vicuna-7B} (left) and \texttt{Llama2-7B} (right).}
    \label{fig:mtbench}
    \vspace{-1.5em}
\end{figure*}

\partitle{Utility evaluation} We summarize the performance of \textit{DELMAN} and baselines on general-purpose tasks with several LLMs on \textit{MT-Bench}, along with seven downstream tasks to comprehensively evaluate the model's utility in Table \ref{tab:utility}. The highest utility scores are highlighted in bold (except \textit{LoRA} which has the highest ASR), and scores that exceed those of the \textit{Original Model} are marked with (\textcolor{green}{$\uparrow$}). 
On \texttt{Vicuna-7B}, \textit{DELMAN} not only slightly outperforms the \textit{Original Model} on \textit{MT-Bench} (6.84 vs 6.77), but also maintains strong performance across \textit{Closed-domain QA}, \textit{NLI}, and \textit{Sentiment analysis}. With \texttt{Llama2-7B}, \textit{DELMAN} demonstrates clear improvements on \textit{NER} (0.228 vs 0.187) and \textit{NLI} (0.612 vs 0.603), and sustains high scores on other downstream tasks. For \texttt{Mistral-7B}, although \textit{DELMAN} is slightly behind the \textit{Original Model} on some metrics, it remains considerably more stable than \textit{SafeDecoding} and \textit{LED}, which both exhibit significant performance drops. On \texttt{Llama3.1-8B}, \textit{DELMAN} surpasses the \textit{Original Model} in \textit{Dialogue}, \textit{NLI}, \textit{Reasoning}, and \textit{Summarization}. For \texttt{Qwen2.5-7B}, \textit{DELMAN} achieves the highest \textit{MT-Bench} score among defenses (8.48), closely matching or exceeding the \textit{Original Model} in multiple downstream tasks. By contrast, other defense methods such as \textit{LED} and \textit{SafeDecoding} may excel in specific tasks but tend to cause substantial overall utility degradation, particularly in comprehensive multi-turn evaluations like \textit{MT-Bench}. In summary, \textit{DELMAN} provides robust defense with minimal utility loss, and in many cases, even improves downstream task performance across various models.
Figures \ref{fig:mtbench} present a detailed breakdown of model performance across \textit{MT-Bench} subcategories. The visualization particularly highlights \textit{DELMAN}'s advantages in preserving complex capabilities, with the largest area marked in dark blue. Notably, \textit{DELMAN} maintains strong performance in Reasoning, Writing, and Roleplay tasks, where \textit{LED} and \textit{SafeDecoding} exhibit substantial weaknesses. This demonstrates \textit{DELMAN}'s ability to balance robustness against jailbreak attacks while minimizing \mbox{degradation in general utility.}


\subsection{Edit According to Harmful Behavior}
In this section, we investigate the effect of \textit{DELMAN} edit on individual harmful behavior and its impact on defending other unedited behavior.

\begin{figure}[]
  \centering
  \setlength{\abovecaptionskip}{-0.2cm}
  \includegraphics[width=\linewidth]{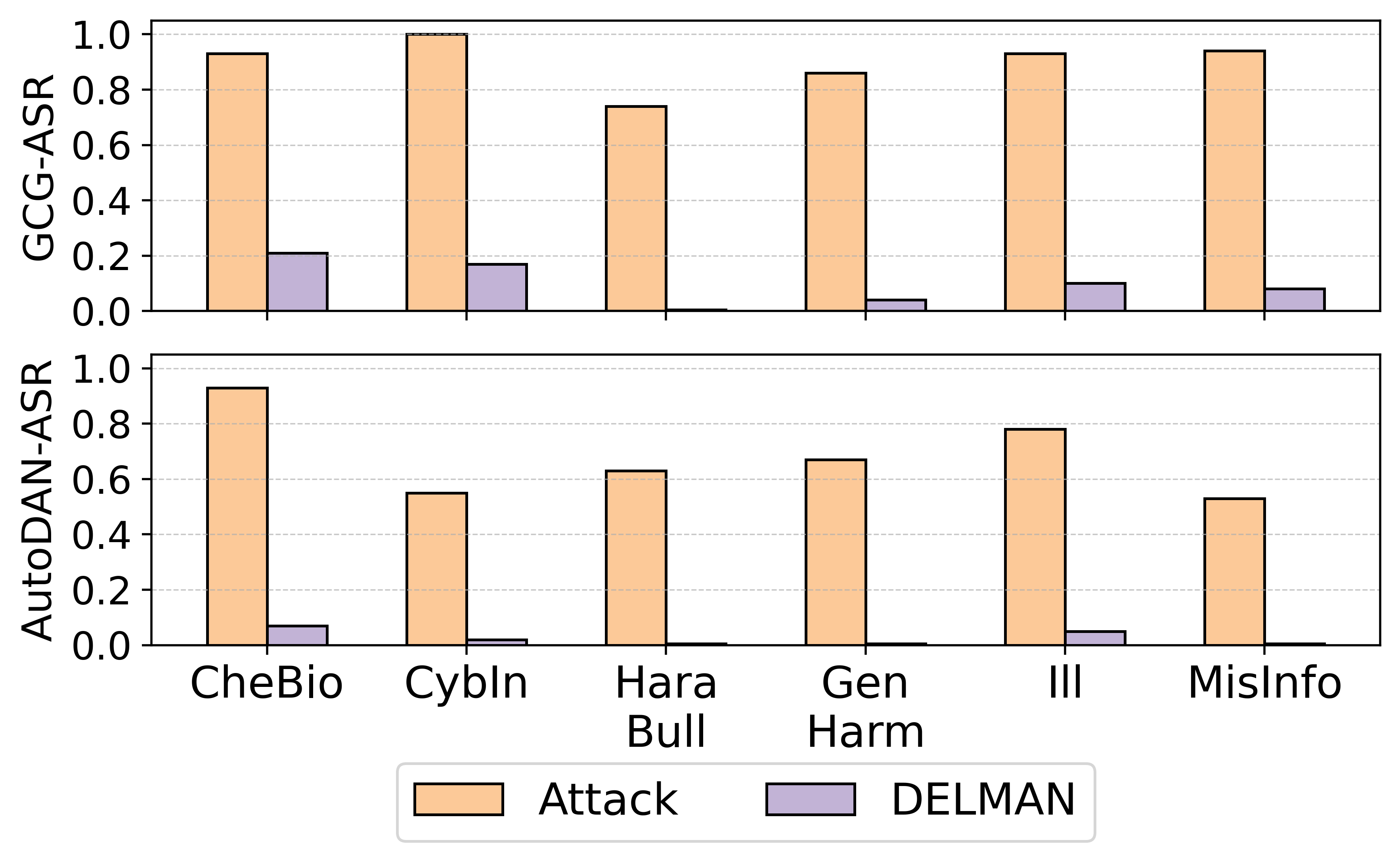}
  \vspace{-.5em}
  \caption{ASR for \texttt{Vicuna-7B} after applying single-behavior \textit{DELMAN} against \textit{GCG} and \textit{AutoDAN} attacks.}
  \label{fig:single-vicuna}
  \vspace{-1.5em} 
\end{figure}

\partitle{Effectiveness of \textit{DELMAN} on each harmful behavior} 
Figure~\ref{fig:single-vicuna} compares the performance of \textit{DELMAN} across individual \textsc{HarmBench} behavior, including chemical and biological (CheBio), cybercrime intrusion (CybIn), harassment and bullying (HaraBull), general harmful (GenHarm), illegal (Ill), and misinformation (MisInfo). The two figures demonstrate the ASR drop on \textit{GCG} and \textit{AutoDAN} after \textit{DELMAN} edits respectively.  In single-behavior editing, \textit{DELMAN} demonstrates significant effectiveness in defending against two types of jailbreak attacks.

\begin{figure}[ht]
    \centering
    \setlength{\abovecaptionskip}{0.1cm} 
    \includegraphics[width=\linewidth]{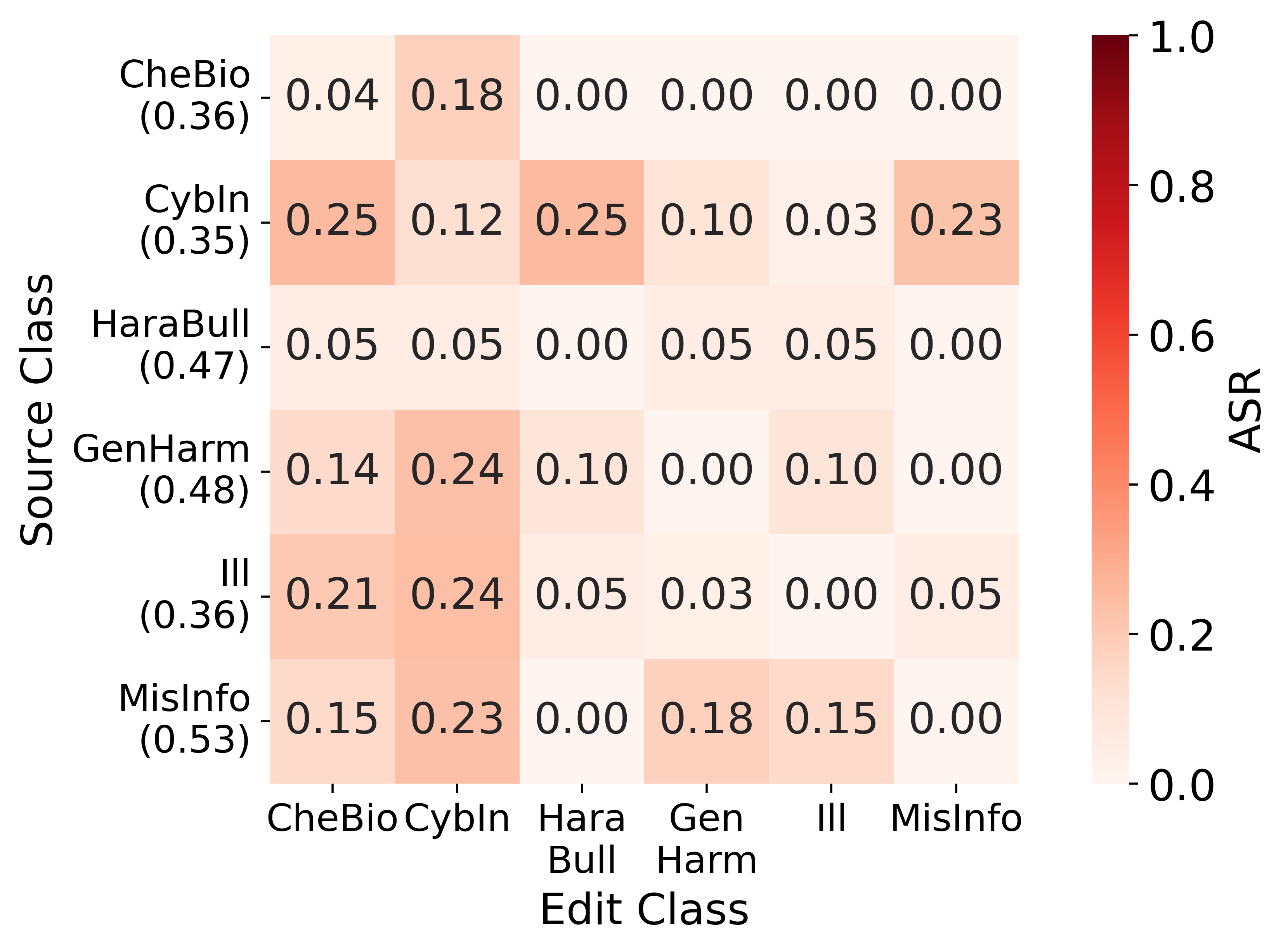}
    \includegraphics[width=\linewidth]{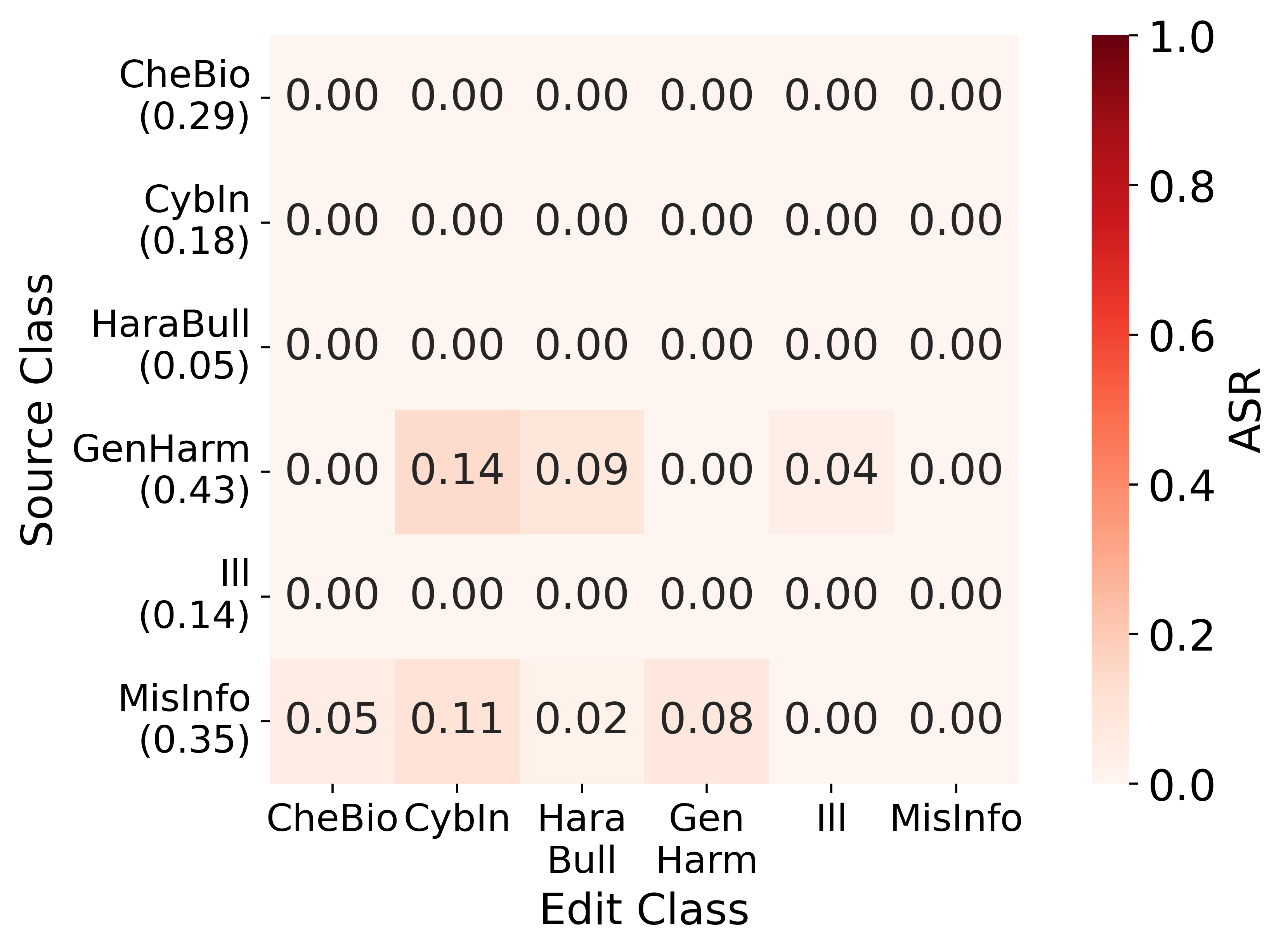}
    \caption{\small
    ASR heatmaps for the cross-behavior  transfer results of single-behavior \textit{DELMAN} edit on \texttt{Llama2-7B} against \textit{GCG} (\textit{up}) and \textit{AutoDAN} (\textit{down}) attacks.}
    \label{fig:llama2_heatmap_gcg_autodan}
    \vspace{-1.5em}
\end{figure}

\begin{figure}[t]
    \centering
    \setlength{\abovecaptionskip}{0.1cm} 
    \begin{subfigure}[t]{0.22\textwidth}
        \centering
        \includegraphics[width=\textwidth]{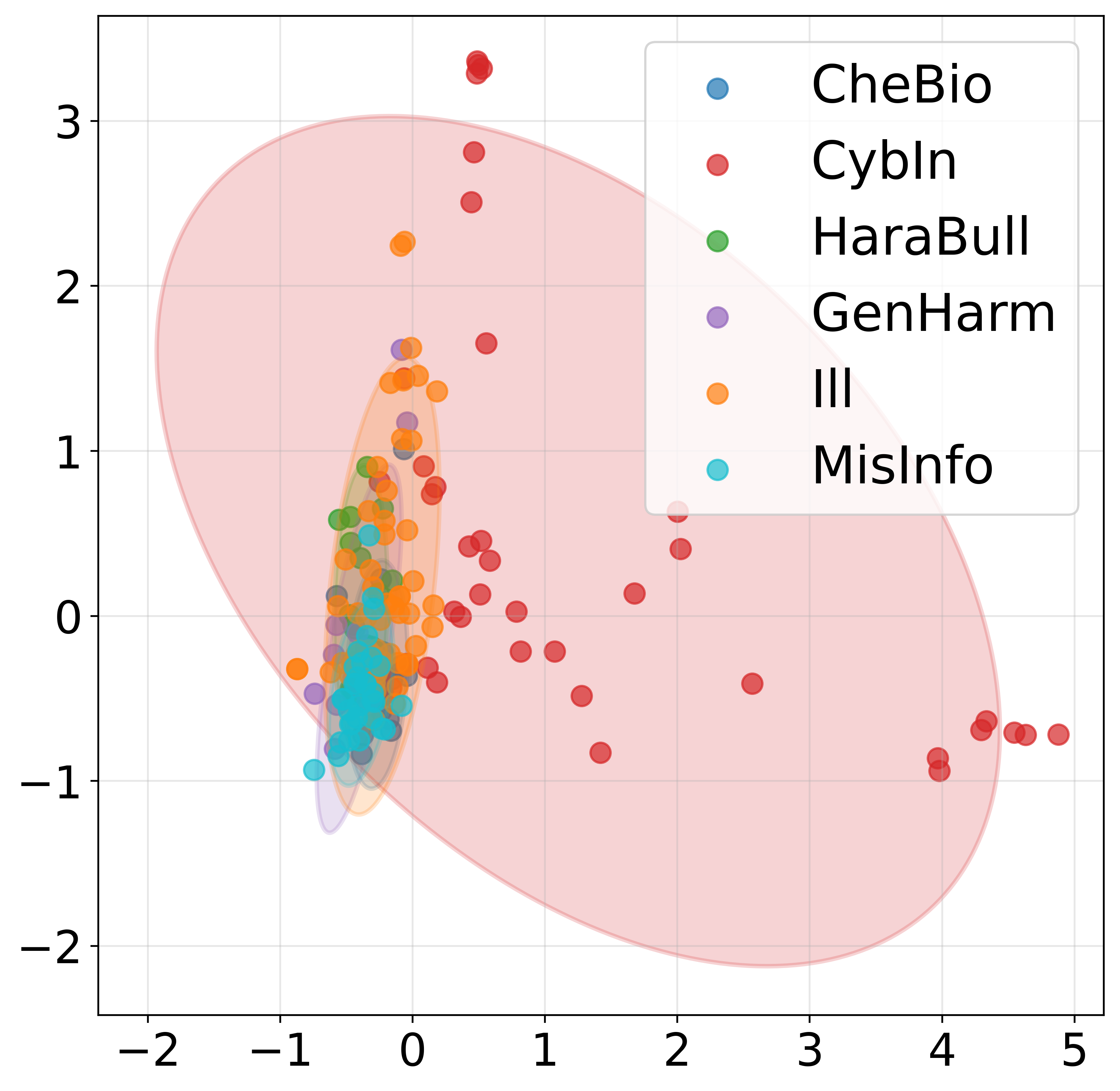} 
        \subcaption{The $k$ of harmful tokens across behaviors.}
        \label{fig:pca1}
    \end{subfigure}
    \hfill
    \begin{subfigure}[t]{0.22\textwidth}
        \centering
        \includegraphics[width=\textwidth]{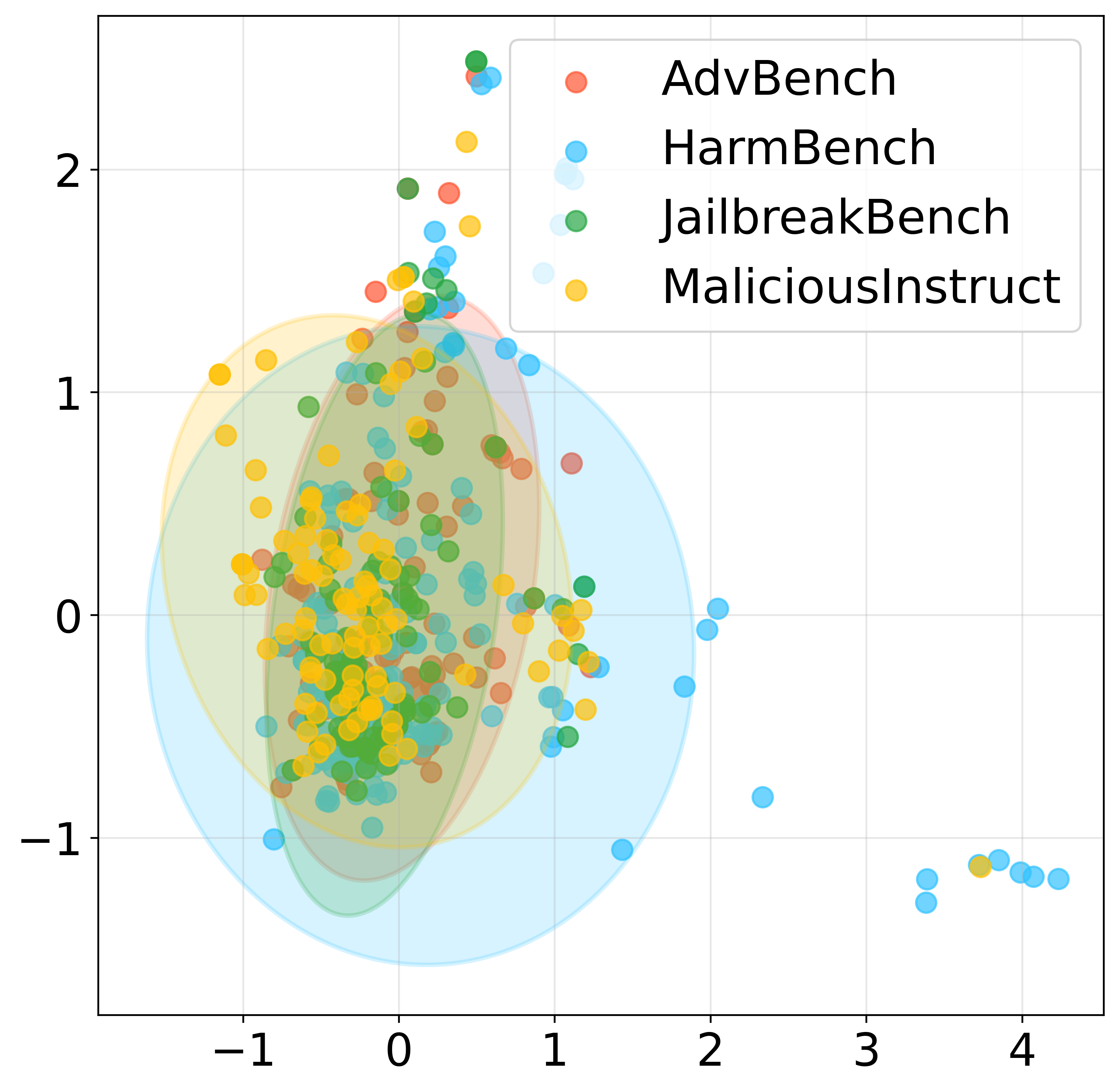} 
        \subcaption{The $k$ of harmful tokens across datasets.}
        \label{fig:pca2}
    \end{subfigure}
    \hfill
    \caption{PCA visualizations of $k$ at the target layer $L$ of \texttt{Llama2-7B} across different behaviors and datasets.}
    \label{fig:two_pca}
    \vspace{-1em}
\end{figure}

\partitle{Cross-behavior observations} We further study the cross-behavior defense performance of \textit{DELMAN} with heatmap. 
We perform single-behavior edits on each behavior with \textit{DELMAN}, and test the resulting model on all six categories, presenting a $6{\times}6$ ASR heatmap. Figure~\ref{fig:llama2_heatmap_gcg_autodan} presents the results for \texttt{Llama2-7B} under the \textit{GCG} and \textit{AutoDAN} jailbreak attacks. Notably, single-category edits in many cases show resilience to off-category attacks. For instance, focusing on CheBio class editing can also mitigate malicious queries from GenHarm or MisInfo classes, reducing ASR even for these distinct domains.

\subsection{Understanding the \textit{DELMAN} Transferability Across Datasets and Behaviors}
\textit{DELMAN} establishes a direct link between harmful tokens and specific responses to modify the model parameters effectively. To explain why modifying the model based on one set of harmful tokens from a specific harmful behavior also improves its robustness against different harmful behavior, and why edits made using examples from one dataset generalize to other datasets, we analyze the distribution of harmful token keys $k$ in the target model layer $l^*$ using Principal Component Analysis (PCA) \cite{PCA}. As shown in Figure \ref{fig:two_pca}, each cluster represents the $k$ of harmful token from a behavior (Figure \ref{fig:pca1}) or from a dataset (Figure \ref{fig:pca2}). We can note that harmful token keys $k$ in the target model layer $l^*$ from different categories or datasets exhibit substantial overlap in the embedding space, suggesting that instructions carrying malicious intent share similar representations across seemingly distinct harm classes or datasets. Through focused editing of these common token representations, \textit{DELMAN} effectively reduces various types of harmful outputs, including those from categories or datasets not seen during editing.


\subsection{Consecutive Edits with \textit{DELMAN}}
\begin{figure}[ht]
    \setlength{\abovecaptionskip}{0.1cm} 
    \centering\includegraphics[width=.8\linewidth]{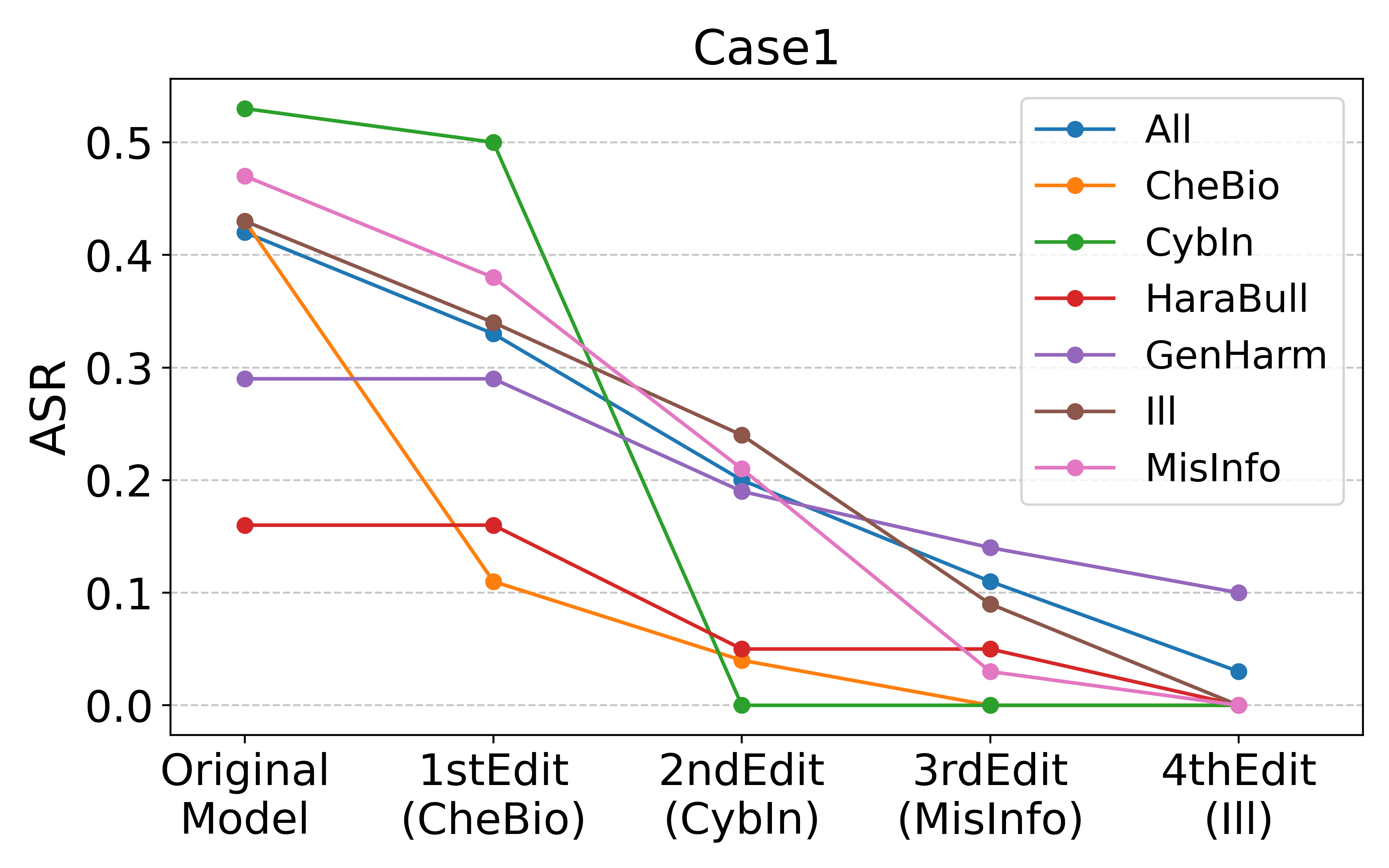}
    \centering\includegraphics[width=.8\linewidth]{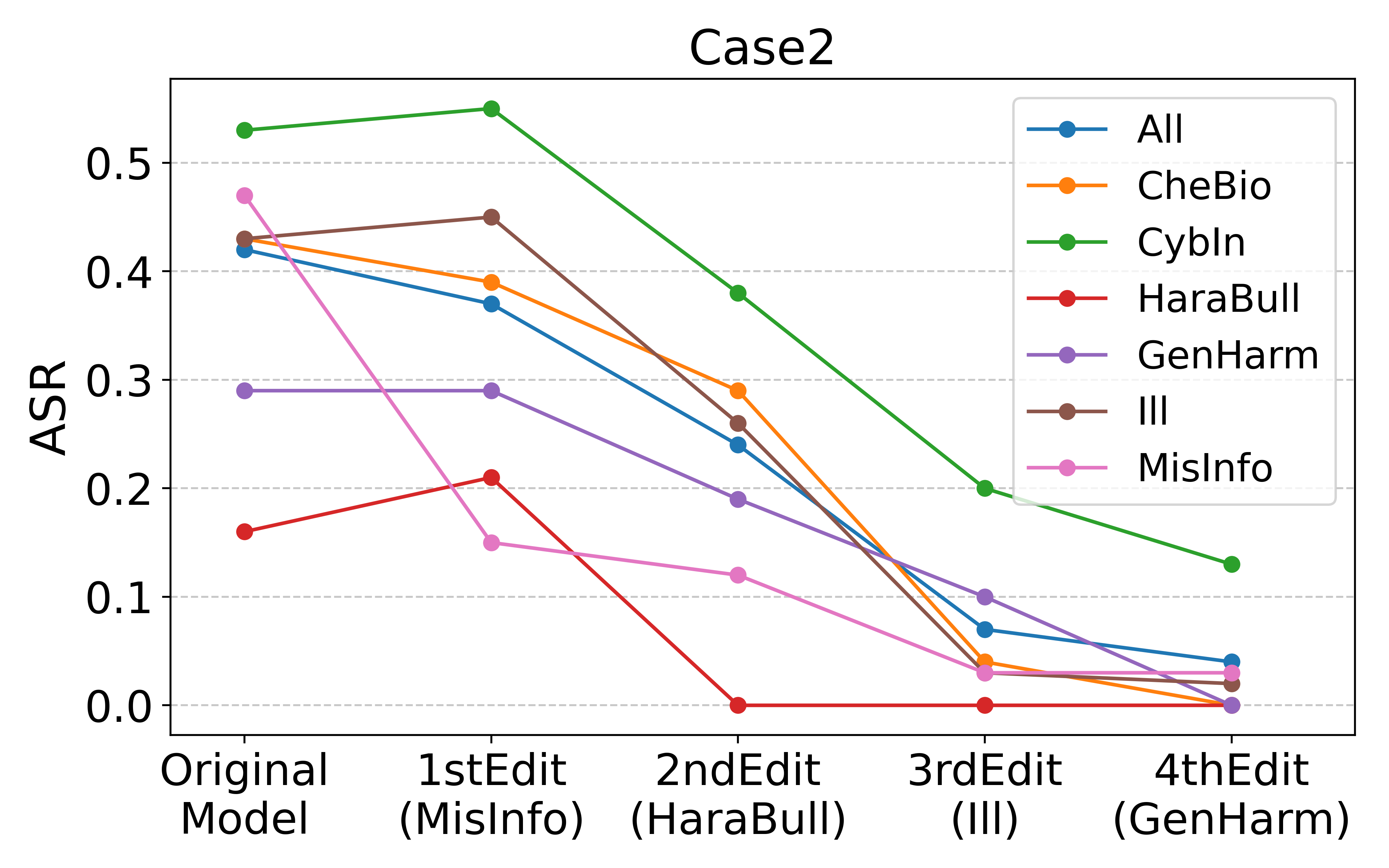}
    \caption{\small
    Defense performance of consecutive \textit{DELMAN} edits on \texttt{Llama2-7B} against \textit{GCG} attacks.}
    \label{fig:case}
    \vspace{-1.5em}
\end{figure}

In real-world deployment, adversarial parties may repeatedly attempt to jailbreak the model, making it crucial for dynamic and consecutive edits to maintain the effects of earlier modifications without interference. To evaluate the robustness of \textit{DELMAN} under consecutive edits, we conduct an experiment where edits are applied sequentially across different harmful behavior categories. Specifically, we select one category each from the HB, AB, JBB, and MI datasets and perform \textit{DELMAN} edits in succession. After each edit, we evaluate:
\begin{itemize}[itemsep=0pt,parsep=0pt,topsep=0pt,partopsep=0pt]
    \item \textbf{ASR on the current edit category} to measure the immediate effectiveness of \textit{DELMAN}.
    \item \textbf{ASR on previously edited categories} to determine whether earlier modifications remain effective.
    \item \textbf{ASR on the full dataset} to assess the overall robustness of \textit{DELMAN} against diverse jailbreak attacks.
\end{itemize}
We used line charts to represent the overall ASR reduction across four successive edit phases for each edited behavior of HB dataset and the ASR of the entire HB dataset. As observed in Figure \ref{fig:case}, the overall ASR for the HB dataset consistently decreases with each edit, indicating that \textit{DELMAN} effectively reduces harmful behaviors across multiple categories and each edit achieves maximal ASR drop in its targeted behavior. Additionally, each category edited during the successive phases maintains its defense effectiveness, with no increase of ASR in subsequent edits. This demonstrates that each edit, once applied, is preserved and does not interfere with the defense applied in previous phases, ensuring continuous and cumulative reduction in \mbox{ASR across the dataset}.


\subsection{Efficiency Evaluation}
We evaluate the efficiency, considering both training time and inference overhead. All results are evaluated on \texttt{Vicuna-7B} with a single A40 GPU, averaged over 5 runs. Table~\ref{tab:efficiency} presents the comparison results across different defense methods.
\textit{DELMAN} achieves optimal performance across all metrics: it requires only 0.4 hours for training (the fastest among all methods), maintains 1× inference overhead unlike \textit{SafeDecoding} (1.07×), and delivers the lowest ASR (6.7\%), demonstrating superior efficiency and effectiveness for practical deployment.
\begin{table}[]
\vspace{-1em}
\centering
\setlength{\abovecaptionskip}{0.2cm} 
\resizebox{0.5\textwidth}{!}{
\begin{tabular}{lccc}
\toprule
\textbf{Method} & \textbf{Training Time} & \textbf{Inference Overhead} & \textbf{Average ASR} \\
\midrule
\textit{LoRA} & 1.5 hr & 1× & 23.2\% \\
\textit{SafeDecoding} & 0.6 hr & 1.07× & 10.7\% \\
\textit{LED} & 6.2 hr & 1× & 8.8\% \\
\textit{DELMAN} &\textbf{0.4 hr} & \textbf{1×} & \textbf{6.7\%} \\
\bottomrule
\end{tabular}
}
\caption{Efficiency comparison of defense methods on \texttt{Vicuna-7B}, \textbf{Bold}: best performance.}
\label{tab:efficiency}
\end{table}

\subsection{Ablation Study}
\vspace{-0.5em}
\partitle{Alternative token identification sources}
To assess the robustness and cost-effectiveness of our approach, we evaluate three different sources for harmful token identification: \texttt{GPT-4} (ours), \texttt{Llama2-7B}, and \texttt{Vicuna-7B}. The results, comparing defense performance and cost across all three sources and averaged over 5 runs, are presented in Table~\ref{tab:token_sources}.
\texttt{GPT-4} achieves the lowest ASR (0.25\%) and highest \textit{MT-Bench} score (6.31), demonstrating that accurate harmful token identification preserves model utility while maintaining strong defense performance. Alternative models offer comparable processing times (8.6-9.5 minutes) but with slightly degraded performance. \texttt{GPT-4} provides the optimal balance of effectiveness and affordability at only \$1.46 token cost, though other LLMs serve as viable alternatives for different deployment scenarios.
\begin{table}[]
\centering
\setlength{\abovecaptionskip}{0.2cm} 
\resizebox{0.5\textwidth}{!}{
\begin{tabular}{lccccc}
\toprule
\textbf{Token} & \textbf{Token Overlap}   & \textbf{Llama2-GCG} & \textbf{MT-Bench} & \textbf{Time Cost} & \textbf{Token} \\
\textbf{Source}& \textbf{with GPT-4 (\%)} & \textbf{ASR (Avg \%)} & \textbf{Score} & \textbf{(min)} & \textbf{Cost}\\
\midrule
GPT-4 (ours) & 100\% & \textbf{0.25\%} & \textbf{6.31} & 9.5 & \$1.46 \\
Llama2-7B & 54.5\% & 1.75\% & 5.39 & \textbf{8.6} & - \\
Vicuna-7B & 58.0\% & 1.5\% & 5.45 & 8.8 & - \\
\bottomrule
\end{tabular}
}
\caption{Comparison of different token identification sources, \textbf{Bold}: best performance.}
\label{tab:token_sources}
\vspace{-0.5em}
\end{table}

\vspace{-1em}
\partitle{Alternative regularization metrics}
We conduct an ablation study examining different regularization metrics: Jensen-Shannon divergence (JS) \cite{lin2002divergence}, Cosine similarity (COS) \cite{salton1975vector}, and KL-divergence (ours) \cite{kullback1951information}. The results are presented in Table~\ref{tab:regularization_metrics}.
KL-divergence consistently achieves the best balance between defense effectiveness and utility preservation. For \texttt{Llama2-7B}, it delivers the lowest ASR while maintaining \textit{MT-Bench} performance (6.31), and for \texttt{Vicuna-7B}, it achieves the highest \textit{MT-Bench} score (6.84) with strong defense. JS divergence preserves utility but shows inconsistent defense, while Cosine similarity provides comparable defense but lower utility scores.
\begin{table}[]
\centering
\setlength{\abovecaptionskip}{0.2cm} 
\resizebox{0.5\textwidth}{!}{
\begin{tabular}{c|cccc|cccc|cccc|c}
\toprule
\multirow{2}{*}{\centering Defense} & \multicolumn{4}{c|}{GCG} & \multicolumn{4}{c|}{AutoDAN} & \multicolumn{4}{c|}{PAIR} & \multirow{2}{*}{\centering MT-Bench}\\
& HB & AB & JBB & MI & HB & AB & JBB & MI & HB & AB & JBB & MI \\
\midrule
\multicolumn{14}{c}{\texttt{Vicuna-7B}} \\
\midrule
\centering \textit{DELMAN\_JS} & 12\%  & 4\%  & 19\%  & 1\%  & 6\%  & 3\%  & 7\%  & 6\%  & 11\%  & 6\%  & \textbf{8\%}  & 5\%  & 6.63\\
\centering \textit{DELMAN\_COS}& 11\%  & 2\%  & 17\%  & 3\%  & 5\%  & 2\% & \textbf{6\%} & 5\% & 10\% & 5\% & 11\% & 5\% & 6.44\\
\centering \textit{DELMAN\_KL} & \textbf{11\%} & \textbf{2\%} & \textbf{17\%} & \textbf{1\%} & \textbf{4\%} & \textbf{2\%} & 8\% & \textbf{5\%} & \textbf{10\%} & \textbf{5\%} & 11\% & \textbf{5\%} & \textbf{6.84}\\
\midrule
\multicolumn{14}{c}{\texttt{Llama2-7B}} \\
\midrule
\centering \textit{DELMAN\_JS} & 1\% & 0\% & 0\% & 1\% & 0\% & 0\% & 0\% & 0\% & 0\% & 0\% & 0\% & 0\% & \textbf{6.33}\\
\centering \textit{DELMAN\_COS} & 1\% & 1\% & 0\% & 1\% & 0\% & 0\% & 0\% & 1\% & 0\% & 0\% & 0\% & 0\% & 6.30\\
\centering \textit{DELMAN\_KL} & \textbf{0\%} & \textbf{0\%} & \textbf{0\%} & \textbf{1\%} & \textbf{0\%} & \textbf{0\%} & \textbf{0\%} & \textbf{0\%} & \textbf{0\%} & \textbf{0\%} & \textbf{0\%} & \textbf{0\%} & 6.31\\
\bottomrule
\end{tabular}
}
\caption{Ablation study on regularization metrics. ASR (\%) across different attacks and datasets. \textbf{Bold}: lowest ASR.}
\label{tab:regularization_metrics}
\vspace{-1.5em}
\end{table}

%% file: 5.Conclusion.tex
\section{Conclusion}
\vspace{-.5em}
In this work, we introduce \textit{DELMAN}, a novel defense mechanism that directly edits model parameters to neutralize harmful behaviors by forming explicit connections. \textit{DELMAN} brings minimal parameter modification, preserving the utility on normal tasks and is capable of dynamic and consecutive edits. Extensive experiments demonstrate superiority over existing baselines in terms of defense performance and utility preservation, as well as strong transferability. Overall, \textit{DELMAN} demonstrates how token-level editing method can effectively enhance model safety while maintaining performance. In the future, it would be interesting to investigate more efficient methods for harmful token identification, for instane, using a minimal set of tokens (e.g., 20-30 Tokens) to effectively cover the majority of harmful scenarios, which would significantly reduce computational costs. Additionally, exploring the application of \textit{DELMAN} to domain-specific LLMs and VLMs would validate its generalizability across different domains and modalities.


\clearpage
\section*{Acknowledgements}
We thank all reviewers for their constructive comments. This work is supported by the Shanghai Engineering Research Center of Intelligent Vision and Imaging and the Open Research Fund of The State Key Laboratory of Blockchain and Data Security, Zhejiang University.

\section*{Limitations}
\noindent The limitations of our study are as follows:

1. Our evaluations are currently restricted to general-purpose LLMs, leaving the applicability to domain-specialized models (e.g., medical or legal LLMs) and larger-scale models (e.g., 70B parameters) unexplored. Further investigation is required to assess its defense capabilities against domain-specific jailbreak attacks and potential impacts on domain expertise after editing.

2. \textit{DELMAN} relies on GPT-4 for harmful token extraction and context generation, which introduces dependency on external models and potential cost barriers.

3. The stability of consecutive edits, though preliminarily validated, needs deeper analysis to assess potential performance drift over extended deployment.

\section*{Ethics Statement}
\textit{DELMAN} directly edits parameters linked to harmful tokens, raising concerns about potential misapplication or unintended bias introduction. We advocate for responsible deployment where practitioners thoroughly validate parameter modifications and strictly limit edits to well-defined harmful content categories. While our approach offers fine-grained, post-deployment protection, it should be viewed as one component within a comprehensive safety framework that includes human oversight and established moderation systems to ensure ethical and harm-free interactions.

%% file: 6.Appendix.tex
\appendix
\section{Algorithm}
\label{app:alg}
Algorithm \ref{alg:DELMAN} demonstrates the detailed procedure of \textit{DELMAN}.

\begin{algorithm*}
\caption{\textbf{\textit{DELMAN}: Dynamic Editing for LLM Jailbreak Defense}}
\label{alg:DELMAN}

\textbf{Input:} Original LLM $f$, Harmful query dataset $\mathcal{Q}_{\text{harm}}$, Target safe response $Y_{\text{target}}$, Target layers $\mathcal{R}$ and the last target layer $L$, Covariance matrix $C^{l}$ for each layer $l \in \mathcal{R}$, Number of random context sequences $N$, KL-divergence factor $\lambda$.

\textbf{Output:} Edited model $f'$

\begin{algorithmic}[1]
\State \textbf{Initialize:} $T_h \gets \varnothing$; $f' \gets f$
\For {$q \in \mathcal{Q}_{harm}$}
    \State $t \gets \operatorname{Extraction}(q)$
\EndFor
\State $T_{h} = \{ t_1, t_2, \dots, t_n \}$
\For{$t \in T_{h}$}
    \For{$j = 1$ to $N$}
        \State $x_{j,t} \gets \operatorname{GenerateSequence}(t)$ 
    \EndFor
\EndFor
\For {$t \in T_{h}$}
    \State ${v^*_t} \gets \underset{{v_t}}{\mathrm{arg\,min}} [L_{safe} + \lambda L_{utility}]$
        \Comment{Eq.\ref{eq:v}}
\EndFor
\State $V_D \gets [v_1^*, v_2^*,\dots, v_n^*]$

\For{$l \in \mathcal{R}$}
    \For {$t \in T_{h}$}
        \For{$j = 1$ to $N$}
        \State $k_{t,j}^{l} \gets \sigma\!\bigl(W_{gate}^{l}\,\gamma(a_{x_j,t}^{l}+h_{x_j,t}^{l-1})\bigr)$
            \Comment{Eq.\ref{eq:k}}
        \EndFor
    \State $k^{l}_t \gets \frac{1}{N}\sum_{j=1}^{N}k_{t,j}^{l}$
        \Comment{Eq.\ref{eq:k}}
    \EndFor
    \State $K^{l}_D \gets [k_1^{l},k_2^{l}, \dots, k_{n}^{l}]$
    \State $R^l_D = \frac{V_D - W_{down}^{L} K_D^L}{L - l + 1}$
         \Comment{Eq.\ref{eq:Rd}}
    \State $f'
    \gets
    W_{down}^{l}
      \;+\;
      R_{D}^{l} \,{K_{D}^{l}}^T
      \bigl(C^{l} \;+\; K_{D}^{l}\,{K_{D}^{l}}^T)^{-1}$
      \Comment{Eq.\ref{eq:W}}
\EndFor
\State return $f'$
\end{algorithmic}
\end{algorithm*}

\section{Detail Experiment Setting}
\subsection{Attack Setup}
\label{app:attack}
For \textit{GCG} \cite{zou2023universal}, \textit{AutoDAN} \cite{liu2023autodan}, and \textit{PAIR} \cite{chao2023jailbreaking}, we leverage multiple harmful query datasets: the complete \textsc{HarmBench} (200 queries), a subset of \textsc{AdvBench} (100 queries), the complete \textsc{JailbreakBench} (100 queries), and the complete \textsc{MaliciousInstruct} (100 queries). For each model under test, we generate specific attack prompts using every query from these datasets, following the original hyperparameter configurations as described in their respective papers.
\subsection{Baseline Setup}
\label{app:baseline}
\subsubsection{Original Model} Recent work \cite{inan2023llama, xu2024comprehensive} has demonstrated that system prompts and prompt formats can significantly influence the effectiveness of jailbreaking attacks. For experimental consistency and reproducibility, we utilize the default settings across all evaluations.

\partitle{Llama2 template}
\begin{tcolorbox}[
    colback=white,
    colframe=black,
    boxrule=0.5pt,
    left=5pt,
    right=5pt,
    top=5pt,
    bottom=5pt,
    breakable=true,
    width=\linewidth,
    fontupper=\small
]
{\ttfamily
[INST] <<SYS>>\\
You are a helpful, respectful and honest assistant. Always answer as helpfully as possible, while being safe. Your answers should not include any harmful, unethical, racist, sexist, toxic, dangerous, or illegal content. Please ensure that your responses are socially unbiased and positive in nature.\\
\\
If a question does not make any sense, or is not factually coherent, explain why instead of answering something not correct. If you don't know the answer to a question, please don't share false information.\\
<</SYS>>\\
\\
\textcolor{blue}{\{instruction\}} [/INST]}
\end{tcolorbox}

\partitle{Vicuna template}
\begin{tcolorbox}[
    colback=white,
    colframe=black,
    boxrule=0.5pt,
    left=5pt,
    right=5pt,
    top=5pt,
    bottom=5pt,
    breakable=true,
    width=\linewidth,
    fontupper=\small
]
{\ttfamily
A chat between a curious human and an artificial intelligence assistant. The assistant gives helpful, detailed, and polite answers to the human's questions. USER: \textcolor{blue}{\{instruction\}} ASSISTANT:}
\end{tcolorbox}

\partitle{Mistral template}
\begin{tcolorbox}[
    colback=white,
    colframe=black,
    boxrule=0.5pt,
    left=5pt,
    right=5pt,
    top=5pt,
    bottom=5pt,
    breakable=true,
    width=\linewidth,
    fontupper=\small
]
{\ttfamily
[INST] \textcolor{blue}{\{instruction\}} [/INST]}
\end{tcolorbox}

\partitle{Llama3.1 template}
\begin{tcolorbox}[
    colback=white,
    colframe=black,
    boxrule=0.5pt,
    left=5pt,
    right=5pt,
    top=5pt,
    bottom=5pt,
    breakable=true,
    width=\linewidth,
    fontupper=\small
]
{\ttfamily
<|begin\_of\_text|><|start\_header\_id|>user\\
<|end\_header\_id|>\\
\\
\textcolor{blue}{\{instruction\}}<|eot\_id|>\\
<|start\_header\_id|>assistant<|end\_header\_id|>\\}
\end{tcolorbox}

\partitle{Qwen2.5 template}
\begin{tcolorbox}[
    colback=white,
    colframe=black,
    boxrule=0.5pt,
    left=5pt,
    right=5pt,
    top=5pt,
    bottom=5pt,
    breakable=true,
    width=\linewidth,
    fontupper=\small
]
{\ttfamily
<|im\_start|>system\\
You are Qwen, created by Alibaba Cloud. You are a helpful assistant.<|im\_end|>\\
<|im\_start|>user\\
\textcolor{blue}{\{instruction\}}<|im\_end|>\\
<|im\_start|>assistant\\}
\end{tcolorbox}

\subsubsection{DELMAN}
\label{app:DELMAN_hp}
The editing dataset consists of 200 queries from \textsc{HarmBench}, along with a single uniform $Y_{target}$: \emph{``I'm sorry, I can't assist with that.''}. Our method does not require a dedicated safe response for every single question; Instead, just this one $Y_{target}$ is sufficient to provide robust defenses with minimal impact on the model’s utility. And we directly adopt the findings of crucial layers in MEMIT \cite{meng2022mass} and EasyEdit \cite{wang2023easyedit} for these models. The hyperparameters of \textit{DELMAN} are shown in Table~\ref{tab:DELMAN_hp}.
\begin{table}[h!]
\centering
\resizebox{0.5\textwidth}{!}{
\begin{tabular}{l|c|c|c|c|c}
\hline
\textbf{Hyperparameter} & \textbf{Llama2} & \textbf{Vicuna} & \textbf{Mistral} & \textbf{Llama3.1} &\textbf{Qwen2.5}\\
\hline
Target Layers $\mathcal{R}$ & [7,8] & [7,8] & [7,8] &[4,5,6,7,8] & [7,8]\\
Learning Rate of $v^*$ & 5e-1 & 5e-1 & 5e-1 & 5e-1 & 5e-1 \\
Weight Decay of $v^*$ & 0.5 & 0.5 & 0.5 & 0.5 & 0.5 \\
Gradient Steps of $v^*$ & 25 & 25 & 25 & 25 & 25\\
Loss Layer of $v^*$ & 31 & 31 & 31 & 31 & 27\\
KL Factor & 0.0625 & 0.0625 & 0.0625 & 0.0625 & 0.0625\\
Clamp Factor & 0.75 & 0.75 & 0.75 & 0.75 & 0.75\\
Mom2 Update Weight & 15000 & 15000 & 15000 & 15000 & 15000\\
Optimizer & Adam & Adam & Adam & Adam & Adam\\
\hline
\end{tabular}
}
\caption{\textit{DELMAN} hyperparameters for different models.}
\label{tab:DELMAN_hp}
\end{table}

\subsubsection{LoRA}
We also apply \textit{LoRA} fine-tuning on the same 200 queries from the \textsc{HarmBench}; However, in this setup, each query is paired with a safe response generated by GPT-4 as the $Y_{target}$. We have verified that these $Y_{target}$ achieve 0 ASR on \textsc{HarmBench} classifier. Notably, if we were to follow the same strategy as used in \textit{DELMAN} and adopt a single uniform $Y_{target}$ for all queries, the model would inevitably converge to generating only that single response. This would severely limit the model's ability to provide diverse and contextually appropriate responses. The hyperparameters of \textit{LoRA} are shown in Table~\ref{tab:lora_hp}.
\begin{table}[h!]
\centering
\resizebox{0.5\textwidth}{!}{
\begin{tabular}{l|c|c|c|c|c}
\hline
\textbf{Hyperparameter} & \textbf{Llama2} & \textbf{Vicuna} & \textbf{Mistral} & \textbf{Llama3.1} &\textbf{Qwen2.5}\\
\hline
LoRA Alpha & 8 & 8 & 8 & 8 & 8\\
LoRA Rank & 32 & 32 & 32 & 32 & 32\\
LoRA Dropout & 0.05 & 0.05 & 0.05 & 0.05 & 0.05\\
Train Batch Size & 1 & 1 & 1 & 1 & 1\\
Grad Accum Steps & 8 & 8 & 8 & 8 & 8\\
Learning Rate & 2e-3 & 2e-3 & 5e-5 & 5e-5 & 5e-5\\
Optimizer & AdamW & AdamW & AdamW & AdamW & AdamW\\
\hline
\end{tabular}
}
\caption{\textit{LoRA} hyperparameters for different models.}
\label{tab:lora_hp}
\end{table}

\subsubsection{SafeDecoding}
SafeDecoding \cite{xu2024safedecoding}, a safety enhancement method that operates by adjusting token probability distributions. This approach strengthens the model's security through two key mechanisms: boosting the probability of safety disclaimers while reducing the likelihood of potential jailbreak sequences. We utilized their publicly released fine-tuned versions of \texttt{Llama2} and \texttt{Vicuna} models, and fine-tuned safer versions of \texttt{Mistral}, \texttt{Llama3.1}, and \texttt{Qwen2.5} using the default training settings.

\subsubsection{LED}
We used the same dataset as in the \textit{LoRA} setup. Since \textit{LED} \cite{zhao2024defending} did not provide an official code implementation, we reproduced their method following the procedures described in their paper. We selected the corresponding layers for each model according to their recommendations. The hyperparameters of \textit{LED} are shown in Table~\ref{tab:LED_hp}.
\begin{table}[h!]
\centering
\resizebox{0.5\textwidth}{!}{
\begin{tabular}{l|c|c|c|c|c}
\hline
\textbf{Hyperparameter} & \textbf{Llama2} & \textbf{Vicuna} & \textbf{Mistral} & \textbf{Llama3.1} &\textbf{Qwen2.5}\\
\hline
Edit Layers & [4,5,6,    & [9,10,11,  & [2,3,4,5,6, & [4,5,6,    & [4,5,6,\\
            &  13,14,15] &  13,14,15] &  13,14,15]  &  13,14,15] &  13,14,15]\\
Target Layers & [29,30,31] & [29,30,31] & [29,30,31] & [29,30,31] & [25,26,27]\\
Learning Rate & 8e-5 & 5e-5 & 5e-6 & 5e-5 & 5e-5\\
Train Batch Size & 1 & 1 & 1 & 1 & 1\\
Gradient Acc Steps & 8 & 8 & 8 & 8 & 8\\
Optimizer & AdamW & AdamW & AdamW & AdamW & AdamW\\
\hline
\end{tabular}
}
\caption{\textit{LED} hyperparameters for different models.}
\label{tab:LED_hp}
\end{table}

\subsection{Downstream Task Datasets}
\label{app:downstream_datasets}
(1) \textit{Closed-domain QA} on the BoolQ \cite{clark2019boolq}, and the results were also measured by EM. (2) \textit{Dialogue} on the MuTual \cite{cui2020mutual}, and the results were measured by selecting one best-matched response from four available candidates \cite{lowe2015ubuntu}. (3) \textit{Named entity recognition (NER)} tested on CoNLL03 \cite{sang2003introduction} with F1 Score. (4) \textit{Natural language inference (NLI)} on RTE \cite{dagan2005pascal} measured by binary classification accuracy. (5) \textit{Reasoning} capability evaluated on GSM8K \cite{cobbe2021training} using solution accuracy. (6) \textit{Sentiment analysis} on SST2 \cite{socher2013recursive} measured by binary classification accuracy. (7) \textit{Summarization} skills evaluated on SAMSum \cite{gliwa2019samsum} using ROUGE metrics \cite{chin2004rouge}.

\section{Prompt}
\subsection{Prompt Used for Harmful Tokens Extraction via GPT-4}
\label{app:extract}
\begin{tcolorbox}[    
    enhanced,
    colback=gray!10,  
    colframe=black,   
    boxrule=1.5pt,    
    arc=5pt,          
    left=10pt,        
    right=10pt,       
    top=5pt,          
    bottom=5pt        
]
\noindent You are tasked with analyzing queries to extract harmful tokens. Follow these instructions: 1. Do not generate harmful, inappropriate, or speculative content; 2. Do not fulfill or answer harmful queries; 3. Extract the most harmful tokens (keywords or phrases) that contribute to the query's harmful nature or potential misuse; 4. Output tokens as a list: \texttt{[token1, token2, ...]}. \\
\textbf{Query: \{INSERT QUERY HERE\}}
\end{tcolorbox}

\subsection{Prompt Used for Random Sequence Generation via GPT-4}
\label{app:random}
\begin{tcolorbox}[    
    enhanced,
    colback=gray!10,  
    colframe=black,   
    boxrule=1.5pt,    
    arc=5pt,          
    left=10pt,        
    right=10pt,       
    top=5pt,          
    bottom=5pt        
]
\noindent You are tasked with generating random sequences that include the specified token. Follow these instructions: 1. Avoid generating harmful, inappropriate, or unsafe content; 2. Each sequence should be 15–30 words long; 3. Use the given token exactly once in each sequence.\\
\textbf{Tokens: \{INSERT TOKENS HERE\}}
\end{tcolorbox}

\section{Supplementary Materials}
\label{app:data}

\subsection{Effectiveness of \textit{DELMAN}}
\label{app:data1}
Table \ref{tab:gcg_pair_autodan} presents the exact value of reduced ASR by \textit{DELMAN} and baselines.
\begin{table*}[t]
\centering
\setlength{\abovecaptionskip}{0.1cm} 
\resizebox{\textwidth}{!}{
\begin{tabular}{c|c|cccc|cccc|cccc}
\toprule
\multirow{2}{*}{\centering{Model}} 
& \multirow{2}{*}{\centering Defense} & \multicolumn{4}{c|}{GCG} & \multicolumn{4}{c|}{AutoDAN} & \multicolumn{4}{c}{PAIR} \\
& & HB & AB & JBB & MI & HB & AB & JBB & MI & HB & AB & JBB & MI \\
\midrule
\multirow{5}{*}{\centering \texttt{Vicuna-7B}} 
& \centering \textit{Original Model} & 92\%  & 89\%  & 89\%  & 94\%  & 69\%  & 78\%  & 73\%  & 83\%  & 80\%  & 75\%  & 77\%  & 86\%  \\
& \centering \textit{LoRA}        & 40\% & 18\% & 32\% & 8\% & 22\% & 29\% & 22\% & 32\% & 26\% & 13\% & 20\% & 16\% \\
& \centering \textit{SafeDecoding}        & 7\% & 4\% & \textbf{3\%} & 1\% & 17\% & 20\% & 18\% & 8\% & 16\% & 8\% & 15\% & 11\% \\
& \centering \textit{LED}        & \textbf{3\%} & 6\% & 34\% & 5\% & 11\% & 9\% & 8\% & 10\% & \textbf{4\%} & 5\% & \textbf{6\%} & 5\% \\
& \centering \textit{DELMAN}     & 11\% & \textbf{2\%} & 17\% & \textbf{1\%} & \textbf{4\%} & \textbf{2\%} & \textbf{8\%} & \textbf{5\%} & 10\% & \textbf{5\% }& 11\% & \textbf{5\%} \\
\midrule
\multirow{5}{*}{\centering \texttt{Llama2-7B}} 
& \centering \textit{Original Model} & 42\% & 39\% & 46\% & 45\% & 23\% & 19\% & 27\% & 30\% & 2\% & 1\% & 4\% & 0\% \\
& \centering \textit{LoRA}        & 13\% & 2\% & 50\% & 32\% & 1\% & 0\% & 1\% & 0\% & 2\% & 0\% & 2\% & 0\% \\
& \centering \textit{SafeDecoding}        & 0\% & 4\% & 1\% & 1\% & 0\% & 0\% & 0\% & 0\% & 1\% & 4\% & 3\% & 0\% \\
& \centering \textit{LED} & 2\% & 0\% & 8\% & 8\% & 2\% & 1\% & 2\% & 2\% & 1\% & 0\% & 4\% & 1\% \\
& \centering \textit{DELMAN} & \textbf{0\%} & \textbf{0\%} & \textbf{0\%} & \textbf{1\%} & \textbf{0\%} & \textbf{0\%} & \textbf{0\%} & \textbf{0\%} & \textbf{0\%} & \textbf{0\%} & \textbf{0\%} & \textbf{0\%} \\
\midrule
\multirow{5}{*}{\centering \texttt{Mistral-7B}} 
& \centering \textit{Original Model} & 80\% & 56\% & 79\% & 94\% & 87\% & 97\% & 95\% & 96\% & 88\% & 82\% & 89\% & 92\% \\
& \centering \textit{LoRA}        & 35\% & 30\% & 38\% & 40\% & 55\% & 60\% & 62\% & 55\% & 42\% & 25\% & 34\% & 44\% \\
& \centering \textit{SafeDecoding}        & 16\% & 12\% & 18\% & 14\% & 15\% & 22\% & 14\% & 24\% & 20\% & 18\% & 22\% & 16\% \\
& \centering \textit{LED} & 20\% & 6\% & 10\% & \textbf{12\%} & 18\% & 26\% & 12\% & 25\% & 19\% & 16\% & \textbf{10\%} & 12\% \\
& \centering \textit{DELMAN} & \textbf{6\%} & \textbf{2\%} & \textbf{2\%} & 14\% & \textbf{8\%} & \textbf{11\%} & \textbf{11\%} & \textbf{12\%} & \textbf{15\%} & \textbf{0\%} & 16\% & \textbf{10\%} \\
\midrule
\multirow{5}{*}{\centering \texttt{Llama3.1-8B}} 
& \centering \textit{Original Model} & 56\% & 47\% & 46\% & 69\% & 20\% & 31\% & 31\% & 46\% & 12\% & 4\% & 11\% & 4\% \\
& \centering \textit{LoRA}        & 41\% & 34\% & 37\% & 42\% & 27\% & 33\% & 35\% & 47\% & 14\% & 4\% & 9\% & 6\% \\
& \centering \textit{SafeDecoding}        & 0\% & 0\% & \textbf{0\%} & 0\% & 4\% & 1\% & 1\% & 1\% & 0\% & 0\% & 1\% & 0\% \\
& \centering \textit{LED} & 1\% & \textbf{0\%} & 3\% & 0\% & 6\% & 7\% & 8\% & 2\% & 0\% & 0\% & 0\% & 1\% \\
& \centering \textit{DELMAN} & \textbf{0\%} & 1\% & 2\% & \textbf{0\%} & \textbf{0\%} & \textbf{0\%} & \textbf{0\%} & \textbf{0\%} & \textbf{0\%} & \textbf{0\%} & \textbf{0\%} & \textbf{0\%} \\
\midrule
\multirow{5}{*}{\centering \texttt{Qwen2.5-7B}} 
& \centering \textit{Original Model} & 49\% & 46\% & 42\% & 80\% & 74\% & 80\% & 80\% & 96\% & 66\% & 39\% & 55\% & 54\% \\
& \centering \textit{LoRA}        & 21\% & 38\% & 36\% & 72\% & 65\% & 82\% & 42\% & 90\% & 32\% & 10\% & 50\% & 44\% \\
& \centering \textit{SafeDecoding}        & 19\% & 17\% & 42\% & 48\% & 77\% & 80\% & 73\% & 97\% & 56\% & 30\% & 43\% & 45\% \\
& \centering \textit{LED} & 1\% & 0\% & 0\% & 0\% & 0\% & 0\% & \textbf{0\%} & 0\% & 5\% & 2\% & \textbf{1\%} & \textbf{2\%} \\
& \centering \textit{DELMAN} & \textbf{0\%} & \textbf{0\%} & \textbf{0\%} & \textbf{0\%} & \textbf{0\%} & \textbf{0\%} & 1\% & \textbf{0\%} & \textbf{3\%} & \textbf{1\%} & 5\% & 4\% \\
\bottomrule
\end{tabular}
}
\caption{ASR (\%) of three jailbreak attacks (\textit{GCG}, \textit{PAIR}, \textit{AutoDAN}) across four datasets on different models, under different defense methods. \textbf{Bold}: lowest ASR.}
\label{tab:gcg_pair_autodan}
\vspace{-1em}
\end{table*}

\subsection{Effectiveness of \textit{DELMAN} on More Models}
\label{app:data2}
Table \ref{tab:extra_models} presents the exact value of reduced ASR by \textit{DELMAN} on additional models.
\begin{table}[h!]
\centering
\setlength{\abovecaptionskip}{0.1cm} 
\resizebox{0.5\textwidth}{!}{
\begin{tabular}{c|c|cccc}
\toprule
\multirow{2}{*}{\centering{Model}} 
& \multirow{2}{*}{\centering Defense} & \multicolumn{4}{c}{GCG}\\
& & HB & AB & JBB & MI\\
\midrule
\multirow{2}{*}{\centering \texttt{Qwen3-8B}} 
& \centering \textit{Original Model} & 63\% & 38\% & 53\% & 90\%\\
& \centering \textit{DELMAN} & \textbf{1\%} & \textbf{0\%} & \textbf{3\%} & \textbf{5\%} \\
\midrule
\multirow{2}{*}{\centering \texttt{Llama-2-13b-chat-hf}} 
& \centering \textit{Original Model} & 14\% & 16\% & 17\% & 18\%\\
& \centering \textit{DELMAN} & \textbf{1\%} & \textbf{0\%} & \textbf{0\%} & \textbf{0\%} \\
\midrule
\multirow{2}{*}{\centering \texttt{Qwen2.5-14B-Instruct}} 
& \centering \textit{Original Model} & 69\% & 42\% & 68\% & 93\%\\
& \centering \textit{DELMAN} & \textbf{3\%} & \textbf{0\%} & \textbf{5\%} & \textbf{3\%} \\
\bottomrule
\end{tabular}
}
\caption{ASR(\%) of \textit{GCG} attack on more LLMs comparing \textit{Original Model} to \textit{DELMAN}. \textbf{Bold}: lowest ASR.}
\label{tab:extra_models}
\vspace{-1em}
\end{table}

\subsection{Effectiveness of \textit{DELMAN} on Each Harmful Behavior}
\begin{figure}[h!]
\setlength{\abovecaptionskip}{0.cm}
  \centering
  \includegraphics[width=\linewidth]{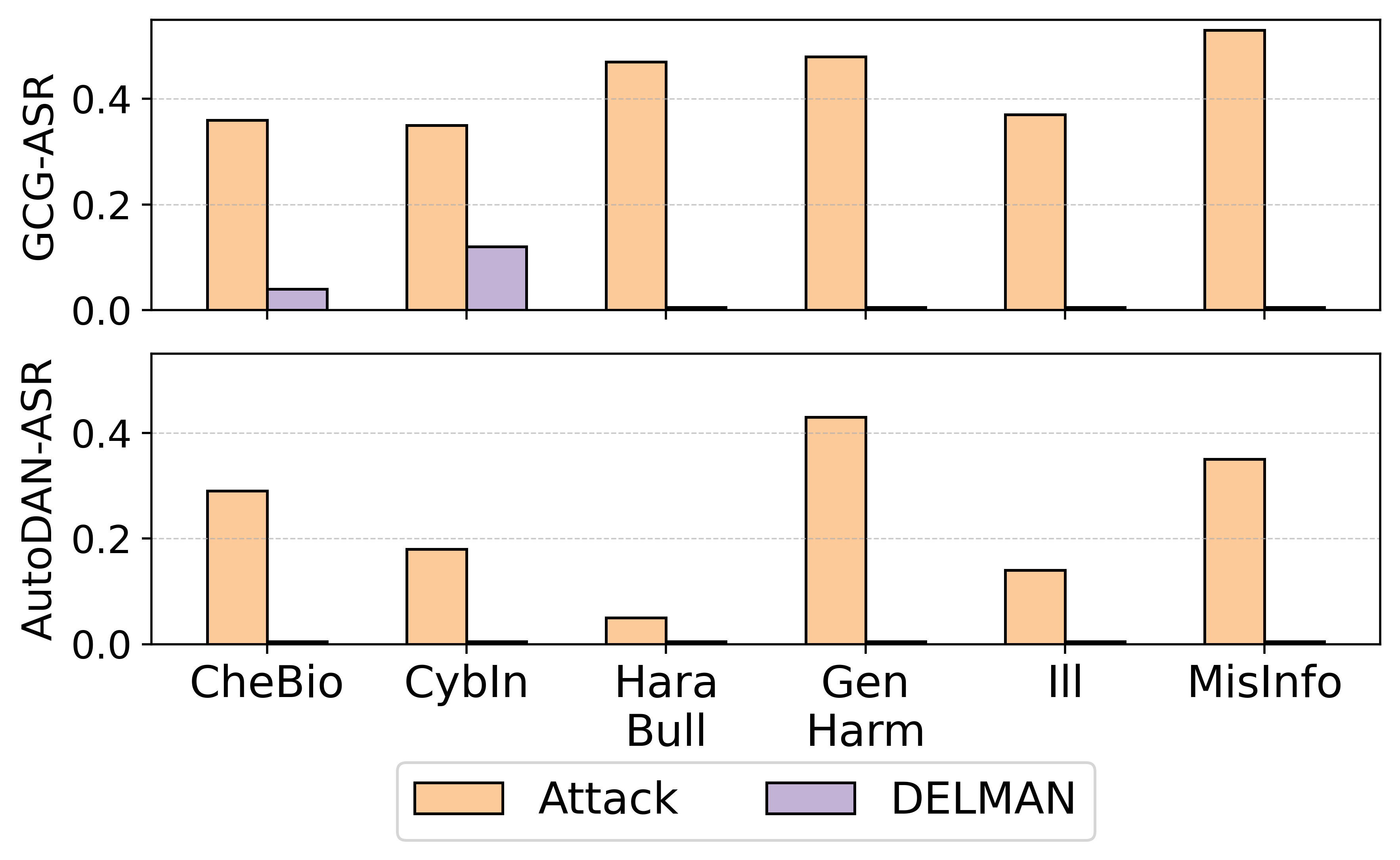}
  \caption{ASR for \texttt{Llama2-7B} after applying single-behavior editing against \textit{GCG} and \textit{AutoDAN} attacks.}
  \label{fig:single-llama}
\end{figure}

Figure~\ref{fig:single-llama} compares the performance of \textit{DELMAN} on \texttt{Llama2-7B} across individual \textsc{HarmBench} behavior.

\subsection{Cross-Behavior Observations}
\begin{figure}[H]
    \centering
    \setlength{\abovecaptionskip}{0.cm}
    \includegraphics[width=\linewidth]{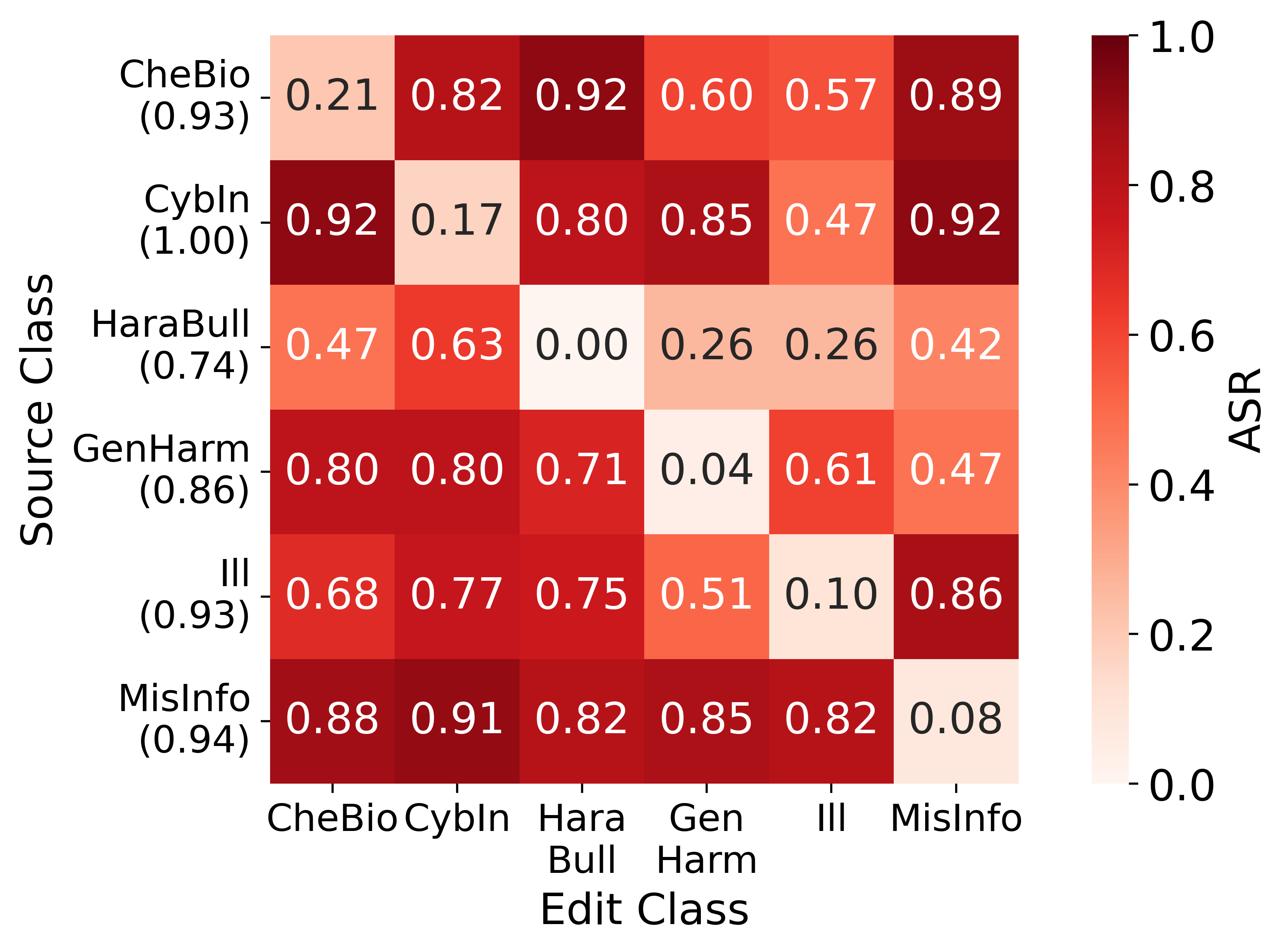}
    \vspace{0.2cm}
    \includegraphics[width=\linewidth]{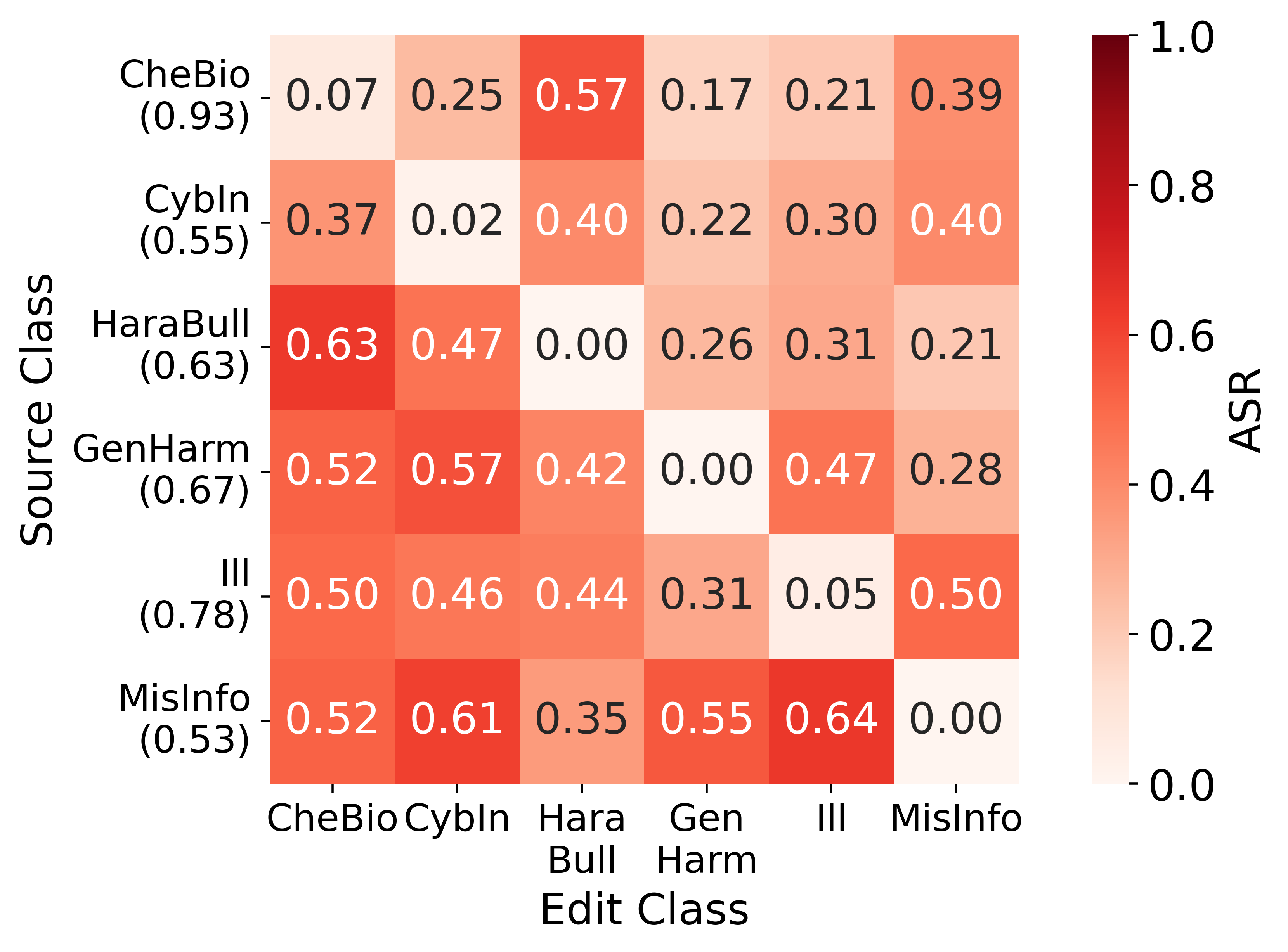}
    \caption{\small
    ASR heatmaps for cross-category transfer results of single-category \textit{DELMAN} defense on \texttt{Vicuna-7B} against \textit{GCG} (top) and \textit{AutoDAN} (bottom) attacks.}
    \label{fig:vicuna_heatmap}
\end{figure}

Figure~\ref{fig:vicuna_heatmap} presents the cross-category transfer results for \texttt{Vicuna-7B} under \textit{GCG} and \textit{AutoDAN} attacks.

\subsection{Effectiveness of Sequential \textit{DELMAN}}
Table \ref{tab:delman_seq} presents experimental results comparing vanilla \textit{DELMAN} with its sequential variants on the \texttt{Llama2-7B} model under GCG attacks.
\begin{table}[h!]
\centering
\resizebox{0.5\textwidth}{!}{
\begin{tabular}{l|c|cccc}
\hline
\multirow{2}{*}{Method} & \multirow{2}{*}{MT-Bench} & \multicolumn{4}{c}{GCG} \\
\cline{3-6}
& & HB & AB & JBB & MI \\
\hline
\textit{DELMAN}                   & 6.31 & \textbf{0\%} & \textbf{0\%} & \textbf{0\%} & \textbf{1\%} \\
\textit{DELMAN(Sequential-Case1)} & 6.35 & 3\% & 0\% & 10\% & 0\% \\
\textit{DELMAN(Sequential-Case2)} & 6.64 & 4\% & 5\% & 6\% & 0\% \\
\hline
\end{tabular}
}
\caption{ASR(\%) of \textit{GCG} attack and \textit{MT-Bench} score on \texttt{Llama2-7B} comparing vanilla \textit{DELMAN} and 4-Edit \textit{DELMAN}. \textbf{Bold}: lowest ASR.}
\label{tab:delman_seq}
\end{table}

\begin{figure}[h!]
    \centering
    \setlength{\abovecaptionskip}{0.1cm} 
    \begin{subfigure}[t]{0.22\textwidth}
        \centering
        \includegraphics[width=\textwidth]{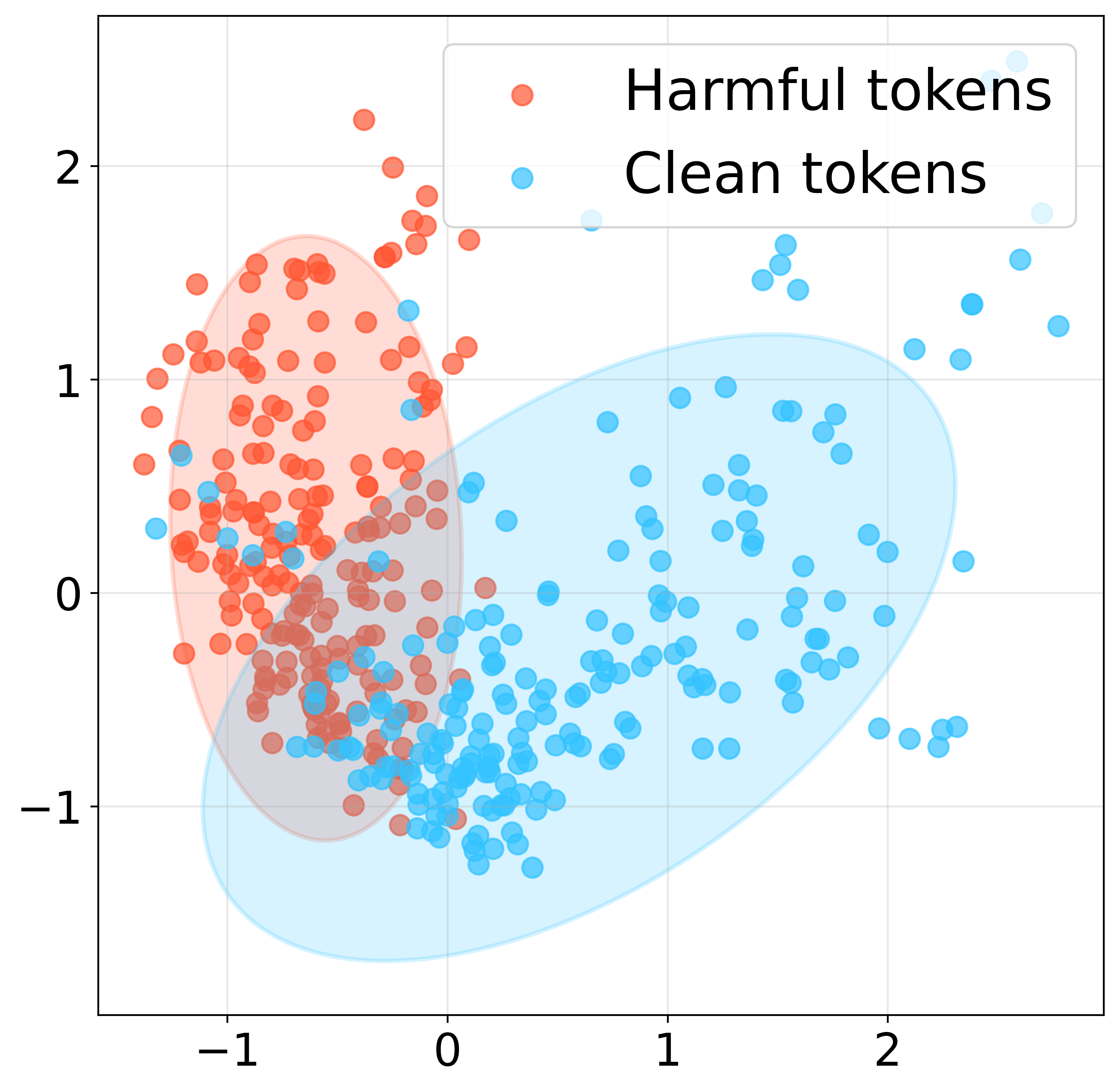} 
        \subcaption{The $k$ of harmful and clean tokens.}
        \label{fig:pca3}
    \end{subfigure}
    \hfill
    \begin{subfigure}[t]{0.22\textwidth}
        \centering
        \includegraphics[width=\textwidth]{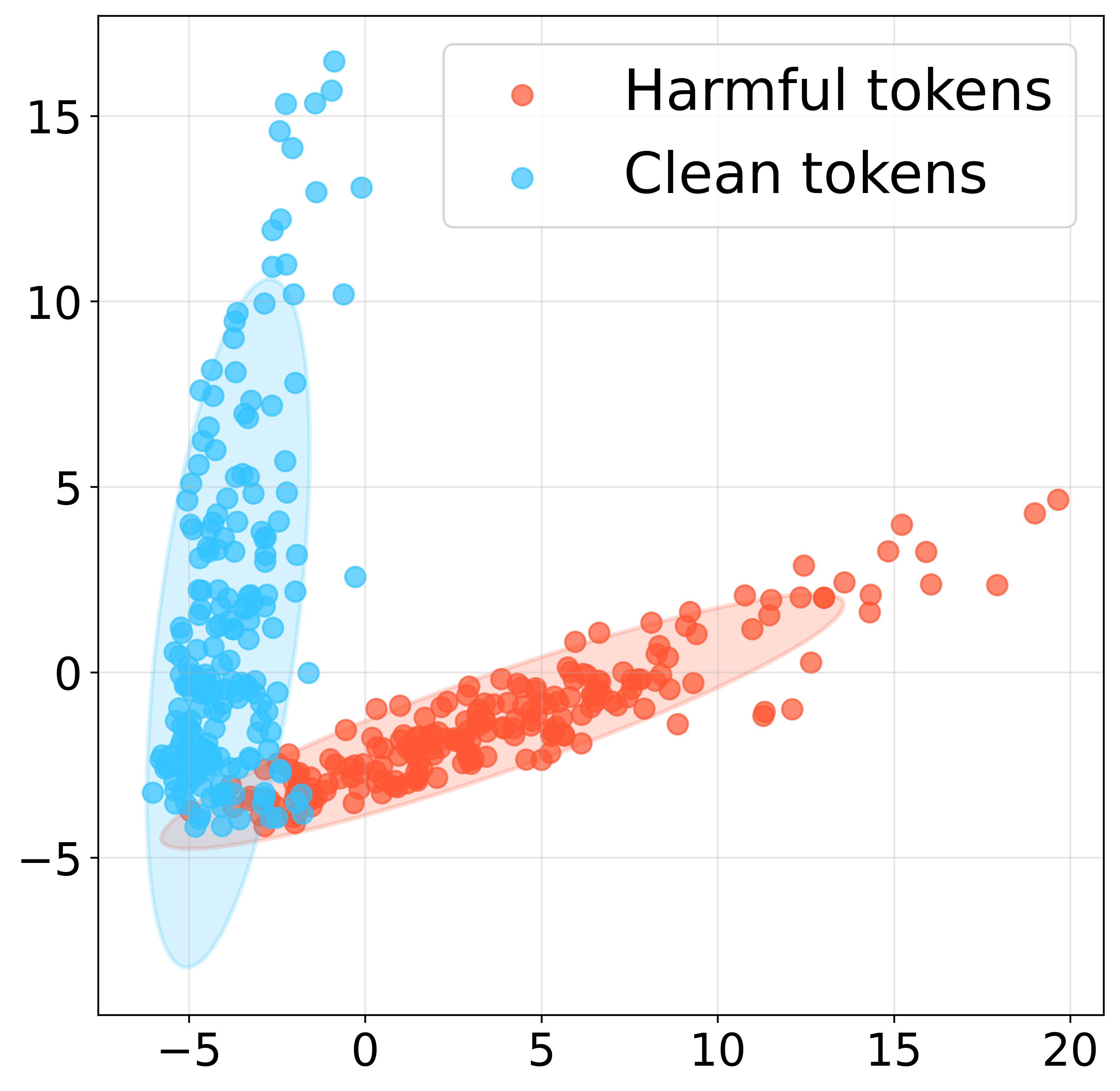} 
        \subcaption{The $v$ of harmful and clean tokens.}
        \label{fig:pca4}
    \end{subfigure}
    \caption{Principal Component Analysis (PCA) visualizations of $k$ and $v$ at the target layer $L$ of \texttt{Llama2-7B} across harmful and clean tokens.}
    \label{fig:harm_clean_pca}
\end{figure}

\subsection{Results of \textit{DELMAN} across Harmful and Clean Tokens}
Figure \ref{fig:harm_clean_pca} shows the $k$ and $v$ distribution differences between harmful and clean tokens. Notably, choosing harmful tokens is vital for preserving model utility: while editing with clean tokens also reduces ASR, these tokens frequently appear in benign queries across various contexts, leading to unnecessary modifications of the model's normal behaviors. In contrast, harmful tokens are primarily concentrated in unsafe queries, allowing for more precise interventions. This explains why editing based on clean tokens leads to significant degradation in \textit{MT-Bench} scores (see Table \ref{tab:delman_harm_clean}) - it unintentionally affects the model's processing of legitimate queries where these common tokens naturally occur. In our experiment, we define clean tokens as the third-to-last word in queries.

\begin{table}[h!]
\centering
\resizebox{0.5\textwidth}{!}{
\begin{tabular}{l|c|cccc}
\hline
\multirow{2}{*}{Method} & \multirow{2}{*}{MT-Bench} & \multicolumn{4}{c}{GCG} \\
\cline{3-6}
& & HB & AB & JBB & MI \\
\hline
\textit{DELMAN} & 6.31 & \textbf{0\%} & \textbf{0\%} & \textbf{0\%} & \textbf{1\%} \\
\textit{DELMAN(clean-token)} & 5.09(\textcolor{red}{$\downarrow$}) & 1\% & 1\% & 3\% & 1\% \\

\hline
\end{tabular}
}
\caption{ASR(\%) of \textit{GCG} attack and \textit{MT-Bench} score on \texttt{Llama2-7B} comparing vanilla \textit{DELMAN} and clean-token \textit{DELMAN}. \textbf{Bold}: lowest ASR.}
\label{tab:delman_harm_clean}
\end{table}

\section{Computing Resources}
The experiments are carried out on 2 NVIDIA A40 GPUs with a total computation time of 1200 GPU hours.

%% file: main.bbl
\begin{thebibliography}{47}
\providecommand{\natexlab}[1]{#1}

\bibitem[{Alon and Kamfonas(2023)}]{alon2023detecting}
Gabriel Alon and Michael Kamfonas. 2023.
\newblock Detecting language model attacks with perplexity.
\newblock \emph{arXiv preprint arXiv:2308.14132}.

\bibitem[{Belinkov and Glass(2019)}]{belinkov2019analysis}
Yonatan Belinkov and James Glass. 2019.
\newblock Analysis methods in neural language processing: A survey.
\newblock \emph{Transactions of the Association for Computational Linguistics}, 7:49--72.

\bibitem[{Cao et~al.(2023)Cao, Cao, Lin, and Chen}]{cao2023defending}
Bochuan Cao, Yuanpu Cao, Lu~Lin, and Jinghui Chen. 2023.
\newblock Defending against alignment-breaking attacks via robustly aligned llm.
\newblock \emph{arXiv preprint arXiv:2309.14348}.

\bibitem[{Chao et~al.(2024)Chao, Debenedetti, Robey, Andriushchenko, Croce, Sehwag, Dobriban, Flammarion, Pappas, Tramer et~al.}]{chao2024jailbreakbench}
Patrick Chao, Edoardo Debenedetti, Alexander Robey, Maksym Andriushchenko, Francesco Croce, Vikash Sehwag, Edgar Dobriban, Nicolas Flammarion, George~J Pappas, Florian Tramer, et~al. 2024.
\newblock Jailbreakbench: An open robustness benchmark for jailbreaking large language models.
\newblock \emph{arXiv preprint arXiv:2404.01318}.

\bibitem[{Chao et~al.(2023)Chao, Robey, Dobriban, Hassani, Pappas, and Wong}]{chao2023jailbreaking}
Patrick Chao, Alexander Robey, Edgar Dobriban, Hamed Hassani, George~J Pappas, and Eric Wong. 2023.
\newblock Jailbreaking black box large language models in twenty queries.
\newblock \emph{arXiv preprint arXiv:2310.08419}.

\bibitem[{Chin-Yew(2004)}]{chin2004rouge}
Lin Chin-Yew. 2004.
\newblock Rouge: A package for automatic evaluation of summaries.
\newblock In \emph{Proceedings of the Workshop on Text Summarization Branches Out, 2004}.

\bibitem[{Clark et~al.(2019)Clark, Lee, Chang, Kwiatkowski, Collins, and Toutanova}]{clark2019boolq}
Christopher Clark, Kenton Lee, Ming-Wei Chang, Tom Kwiatkowski, Michael Collins, and Kristina Toutanova. 2019.
\newblock Boolq: Exploring the surprising difficulty of natural yes/no questions.
\newblock \emph{arXiv preprint arXiv:1905.10044}.

\bibitem[{Cobbe et~al.(2021)Cobbe, Kosaraju, Bavarian, Chen, Jun, Kaiser, Plappert, Tworek, Hilton, Nakano et~al.}]{cobbe2021training}
Karl Cobbe, Vineet Kosaraju, Mohammad Bavarian, Mark Chen, Heewoo Jun, Lukasz Kaiser, Matthias Plappert, Jerry Tworek, Jacob Hilton, Reiichiro Nakano, et~al. 2021.
\newblock Training verifiers to solve math word problems.
\newblock \emph{arXiv preprint arXiv:2110.14168}.

\bibitem[{Cui et~al.(2020)Cui, Wu, Liu, Zhang, and Zhou}]{cui2020mutual}
Leyang Cui, Yu~Wu, Shujie Liu, Yue Zhang, and Ming Zhou. 2020.
\newblock Mutual: A dataset for multi-turn dialogue reasoning.
\newblock \emph{arXiv preprint arXiv:2004.04494}.

\bibitem[{Dagan et~al.(2005)Dagan, Glickman, and Magnini}]{dagan2005pascal}
Ido Dagan, Oren Glickman, and Bernardo Magnini. 2005.
\newblock The pascal recognising textual entailment challenge.
\newblock In \emph{Machine learning challenges workshop}, pages 177--190. Springer.

\bibitem[{De~Cao et~al.(2021)De~Cao, Aziz, and Titov}]{de2021editing}
Nicola De~Cao, Wilker Aziz, and Ivan Titov. 2021.
\newblock Editing factual knowledge in language models.
\newblock \emph{arXiv preprint arXiv:2104.08164}.

\bibitem[{Ganguli et~al.(2022)Ganguli, Lovitt, Kernion, Askell, Bai, Kadavath, Mann, Perez, Schiefer, Ndousse et~al.}]{ganguli2022red}
Deep Ganguli, Liane Lovitt, Jackson Kernion, Amanda Askell, Yuntao Bai, Saurav Kadavath, Ben Mann, Ethan Perez, Nicholas Schiefer, Kamal Ndousse, et~al. 2022.
\newblock Red teaming language models to reduce harms: Methods, scaling behaviors, and lessons learned.
\newblock \emph{arXiv preprint arXiv:2209.07858}.

\bibitem[{Geva et~al.(2022)Geva, Caciularu, Wang, and Goldberg}]{geva2022transformer}
Mor Geva, Avi Caciularu, Kevin~Ro Wang, and Yoav Goldberg. 2022.
\newblock Transformer feed-forward layers build predictions by promoting concepts in the vocabulary space.
\newblock \emph{arXiv preprint arXiv:2203.14680}.

\bibitem[{Geva et~al.(2020)Geva, Schuster, Berant, and Levy}]{geva2020transformer}
Mor Geva, Roei Schuster, Jonathan Berant, and Omer Levy. 2020.
\newblock Transformer feed-forward layers are key-value memories.
\newblock \emph{arXiv preprint arXiv:2012.14913}.

\bibitem[{Gliwa et~al.(2019)Gliwa, Mochol, Biesek, and Wawer}]{gliwa2019samsum}
Bogdan Gliwa, Iwona Mochol, Maciej Biesek, and Aleksander Wawer. 2019.
\newblock Samsum corpus: A human-annotated dialogue dataset for abstractive summarization.
\newblock \emph{arXiv preprint arXiv:1911.12237}.

\bibitem[{Grattafiori et~al.(2024)Grattafiori, Dubey, Jauhri, Pandey, Kadian, Al-Dahle, Letman, Mathur, Schelten, Vaughan et~al.}]{grattafiori2024llama}
Aaron Grattafiori, Abhimanyu Dubey, Abhinav Jauhri, Abhinav Pandey, Abhishek Kadian, Ahmad Al-Dahle, Aiesha Letman, Akhil Mathur, Alan Schelten, Alex Vaughan, et~al. 2024.
\newblock The llama 3 herd of models.
\newblock \emph{arXiv preprint arXiv:2407.21783}.

\bibitem[{Hu et~al.(2021)Hu, Shen, Wallis, Allen-Zhu, Li, Wang, Wang, and Chen}]{hu2021lora}
Edward~J Hu, Yelong Shen, Phillip Wallis, Zeyuan Allen-Zhu, Yuanzhi Li, Shean Wang, Lu~Wang, and Weizhu Chen. 2021.
\newblock Lora: Low-rank adaptation of large language models.
\newblock \emph{arXiv preprint arXiv:2106.09685}.

\bibitem[{Huang et~al.(2023)Huang, Gupta, Xia, Li, and Chen}]{huang2023catastrophic}
Yangsibo Huang, Samyak Gupta, Mengzhou Xia, Kai Li, and Danqi Chen. 2023.
\newblock Catastrophic jailbreak of open-source llms via exploiting generation.
\newblock \emph{arXiv preprint arXiv:2310.06987}.

\bibitem[{Inan et~al.(2023)Inan, Upasani, Chi, Rungta, Iyer, Mao, Tontchev, Hu, Fuller, Testuggine et~al.}]{inan2023llama}
Hakan Inan, Kartikeya Upasani, Jianfeng Chi, Rashi Rungta, Krithika Iyer, Yuning Mao, Michael Tontchev, Qing Hu, Brian Fuller, Davide Testuggine, et~al. 2023.
\newblock Llama guard: Llm-based input-output safeguard for human-ai conversations.
\newblock \emph{arXiv preprint arXiv:2312.06674}.

\bibitem[{Jain et~al.(2023)Jain, Schwarzschild, Wen, Somepalli, Kirchenbauer, Chiang, Goldblum, Saha, Geiping, and Goldstein}]{jain2023baseline}
Neel Jain, Avi Schwarzschild, Yuxin Wen, Gowthami Somepalli, John Kirchenbauer, Ping-yeh Chiang, Micah Goldblum, Aniruddha Saha, Jonas Geiping, and Tom Goldstein. 2023.
\newblock Baseline defenses for adversarial attacks against aligned language models.
\newblock \emph{arXiv preprint arXiv:2309.00614}.

\bibitem[{Jiang et~al.(2024)Jiang, Sablayrolles, Mensch, Bamford, Chaplot, Casas, Bressand, Lengyel, Lample, Saulnier et~al.}]{jiang2024mistral}
AQ~Jiang, A~Sablayrolles, A~Mensch, C~Bamford, DS~Chaplot, Ddl Casas, F~Bressand, G~Lengyel, G~Lample, L~Saulnier, et~al. 2024.
\newblock Mistral 7b. arxiv 2023.
\newblock \emph{arXiv preprint arXiv:2310.06825}.

\bibitem[{Kullback and Leibler(1951)}]{kullback1951information}
Solomon Kullback and Richard~A Leibler. 1951.
\newblock On information and sufficiency.
\newblock \emph{The annals of mathematical statistics}, 22(1):79--86.

\bibitem[{Lee et~al.(2022)Lee, Han, Hwang, Lee, Park, and Lee}]{lee2022plug}
Kyungjae Lee, Wookje Han, Seung-won Hwang, Hwaran Lee, Joonsuk Park, and Sang-Woo Lee. 2022.
\newblock Plug-and-play adaptation for continuously-updated qa.
\newblock \emph{arXiv preprint arXiv:2204.12785}.

\bibitem[{Lin(2002)}]{lin2002divergence}
Jianhua Lin. 2002.
\newblock Divergence measures based on the shannon entropy.
\newblock \emph{IEEE Transactions on Information theory}, 37(1):145--151.

\bibitem[{Liu et~al.(2023)Liu, Xu, Chen, and Xiao}]{liu2023autodan}
Xiaogeng Liu, Nan Xu, Muhao Chen, and Chaowei Xiao. 2023.
\newblock Autodan: Generating stealthy jailbreak prompts on aligned large language models.
\newblock \emph{arXiv preprint arXiv:2310.04451}.

\bibitem[{Lowe et~al.(2015)Lowe, Pow, Serban, and Pineau}]{lowe2015ubuntu}
Ryan Lowe, Nissan Pow, Iulian Serban, and Joelle Pineau. 2015.
\newblock The ubuntu dialogue corpus: A large dataset for research in unstructured multi-turn dialogue systems.
\newblock \emph{arXiv preprint arXiv:1506.08909}.

\bibitem[{Mazeika et~al.(2024)Mazeika, Phan, Yin, Zou, Wang, Mu, Sakhaee, Li, Basart, Li et~al.}]{mazeika2024harmbench}
Mantas Mazeika, Long Phan, Xuwang Yin, Andy Zou, Zifan Wang, Norman Mu, Elham Sakhaee, Nathaniel Li, Steven Basart, Bo~Li, et~al. 2024.
\newblock Harmbench: A standardized evaluation framework for automated red teaming and robust refusal.
\newblock \emph{arXiv preprint arXiv:2402.04249}.

\bibitem[{Meng et~al.(2022{\natexlab{a}})Meng, Bau, Andonian, and Belinkov}]{meng2022locating}
Kevin Meng, David Bau, Alex Andonian, and Yonatan Belinkov. 2022{\natexlab{a}}.
\newblock Locating and editing factual associations in gpt.
\newblock \emph{Advances in Neural Information Processing Systems}, 35:17359--17372.

\bibitem[{Meng et~al.(2022{\natexlab{b}})Meng, Sharma, Andonian, Belinkov, and Bau}]{meng2022mass}
Kevin Meng, Arnab~Sen Sharma, Alex Andonian, Yonatan Belinkov, and David Bau. 2022{\natexlab{b}}.
\newblock Mass-editing memory in a transformer.
\newblock \emph{arXiv preprint arXiv:2210.07229}.

\bibitem[{Mitchell et~al.(2021)Mitchell, Lin, Bosselut, Finn, and Manning}]{mitchell2021fast}
Eric Mitchell, Charles Lin, Antoine Bosselut, Chelsea Finn, and Christopher~D Manning. 2021.
\newblock Fast model editing at scale.
\newblock \emph{arXiv preprint arXiv:2110.11309}.

\bibitem[{Salton et~al.(1975)Salton, Wong, and Yang}]{salton1975vector}
Gerard Salton, Anita Wong, and Chung-Shu Yang. 1975.
\newblock A vector space model for automatic indexing.
\newblock \emph{Communications of the ACM}, 18(11):613--620.

\bibitem[{Sang and De~Meulder(2003)}]{sang2003introduction}
Erik~F Sang and Fien De~Meulder. 2003.
\newblock Introduction to the conll-2003 shared task: Language-independent named entity recognition.
\newblock \emph{arXiv preprint cs/0306050}.

\bibitem[{Socher et~al.(2013)Socher, Perelygin, Wu, Chuang, Manning, Ng, and Potts}]{socher2013recursive}
Richard Socher, Alex Perelygin, Jean Wu, Jason Chuang, Christopher~D Manning, Andrew~Y Ng, and Christopher Potts. 2013.
\newblock Recursive deep models for semantic compositionality over a sentiment treebank.
\newblock In \emph{Proceedings of the 2013 conference on empirical methods in natural language processing}, pages 1631--1642.

\bibitem[{Touvron et~al.(2023)Touvron, Martin, Stone, Albert, Almahairi, Babaei, Bashlykov, Batra, Bhargava, Bhosale et~al.}]{touvron2023llama}
Hugo Touvron, Louis Martin, Kevin Stone, Peter Albert, Amjad Almahairi, Yasmine Babaei, Nikolay Bashlykov, Soumya Batra, Prajjwal Bhargava, Shruti Bhosale, et~al. 2023.
\newblock Llama 2: Open foundation and fine-tuned chat models.
\newblock \emph{arXiv preprint arXiv:2307.09288}.

\bibitem[{Wang et~al.(2024)Wang, Zhang, Xu, Xi, Deng, Yao, Zhang, Yang, Wang, and Chen}]{wang2024detoxifying}
Mengru Wang, Ningyu Zhang, Ziwen Xu, Zekun Xi, Shumin Deng, Yunzhi Yao, Qishen Zhang, Linyi Yang, Jindong Wang, and Huajun Chen. 2024.
\newblock Detoxifying large language models via knowledge editing.
\newblock \emph{arXiv preprint arXiv:2403.14472}.

\bibitem[{Wang et~al.(2023)Wang, Zhang, Tian, Xi, Yao, Xu, Wang, Mao, Wang, Cheng et~al.}]{wang2023easyedit}
Peng Wang, Ningyu Zhang, Bozhong Tian, Zekun Xi, Yunzhi Yao, Ziwen Xu, Mengru Wang, Shengyu Mao, Xiaohan Wang, Siyuan Cheng, et~al. 2023.
\newblock Easyedit: An easy-to-use knowledge editing framework for large language models.
\newblock \emph{arXiv preprint arXiv:2308.07269}.

\bibitem[{Wang et~al.(2022)Wang, Kordi, Mishra, Liu, Smith, Khashabi, and Hajishirzi}]{wang2022self}
Yizhong Wang, Yeganeh Kordi, Swaroop Mishra, Alisa Liu, Noah~A Smith, Daniel Khashabi, and Hannaneh Hajishirzi. 2022.
\newblock Self-instruct: Aligning language models with self-generated instructions.
\newblock \emph{arXiv preprint arXiv:2212.10560}.

\bibitem[{Wold et~al.(1987)Wold, Esbensen, and Geladi}]{PCA}
Svante Wold, Kim Esbensen, and Paul Geladi. 1987.
\newblock Principal component analysis.
\newblock \emph{Chemometrics and intelligent laboratory systems}, 2(1-3):37--52.

\bibitem[{Xu et~al.(2024{\natexlab{a}})Xu, Jiang, Niu, Jia, Lin, and Poovendran}]{xu2024safedecoding}
Zhangchen Xu, Fengqing Jiang, Luyao Niu, Jinyuan Jia, Bill~Yuchen Lin, and Radha Poovendran. 2024{\natexlab{a}}.
\newblock Safedecoding: Defending against jailbreak attacks via safety-aware decoding.
\newblock \emph{arXiv preprint arXiv:2402.08983}.

\bibitem[{Xu et~al.(2024{\natexlab{b}})Xu, Liu, Deng, Li, and Picek}]{xu2024comprehensive}
Zihao Xu, Yi~Liu, Gelei Deng, Yuekang Li, and Stjepan Picek. 2024{\natexlab{b}}.
\newblock A comprehensive study of jailbreak attack versus defense for large language models.
\newblock In \emph{Findings of the Association for Computational Linguistics ACL 2024}, pages 7432--7449.

\bibitem[{Yang et~al.(2024)Yang, Yang, Zhang, Hui, Zheng, Yu, Li, Liu, Huang, Wei, Lin, Yang, Tu, Zhang, Yang, Yang, Zhou, Lin, Dang, Lu, Bao, Yang, Yu, Li, Xue, Zhang, Zhu, Men, Lin, Li, Xia, Ren, Ren, Fan, Su, Zhang, Wan, Liu, Cui, Zhang, and Qiu}]{qwen2.5}
An~Yang, Baosong Yang, Beichen Zhang, Binyuan Hui, Bo~Zheng, Bowen Yu, Chengyuan Li, Dayiheng Liu, Fei Huang, Haoran Wei, Huan Lin, Jian Yang, Jianhong Tu, Jianwei Zhang, Jianxin Yang, Jiaxi Yang, Jingren Zhou, Junyang Lin, Kai Dang, Keming Lu, Keqin Bao, Kexin Yang, Le~Yu, Mei Li, Mingfeng Xue, Pei Zhang, Qin Zhu, Rui Men, Runji Lin, Tianhao Li, Tingyu Xia, Xingzhang Ren, Xuancheng Ren, Yang Fan, Yang Su, Yichang Zhang, Yu~Wan, Yuqiong Liu, Zeyu Cui, Zhenru Zhang, and Zihan Qiu. 2024.
\newblock Qwen2.5 technical report.
\newblock \emph{arXiv preprint arXiv:2412.15115}.

\bibitem[{Zhao et~al.(2024)Zhao, Li, Li, Zhang, and Sun}]{zhao2024defending}
Wei Zhao, Zhe Li, Yige Li, Ye~Zhang, and Jun Sun. 2024.
\newblock Defending large language models against jailbreak attacks via layer-specific editing.
\newblock \emph{arXiv preprint arXiv:2405.18166}.

\bibitem[{Zheng et~al.(2023)Zheng, Chiang, Sheng, Zhuang, Wu, Zhuang, Lin, Li, Li, Xing et~al.}]{zheng2023judging}
Lianmin Zheng, Wei-Lin Chiang, Ying Sheng, Siyuan Zhuang, Zhanghao Wu, Yonghao Zhuang, Zi~Lin, Zhuohan Li, Dacheng Li, Eric Xing, et~al. 2023.
\newblock Judging llm-as-a-judge with mt-bench and chatbot arena.
\newblock \emph{Advances in Neural Information Processing Systems}, 36:46595--46623.

\bibitem[{Zhou et~al.(2024{\natexlab{a}})Zhou, Li, and Wang}]{zhou2024robust}
Andy Zhou, Bo~Li, and Haohan Wang. 2024{\natexlab{a}}.
\newblock Robust prompt optimization for defending language models against jailbreaking attacks.
\newblock \emph{arXiv preprint arXiv:2401.17263}.

\bibitem[{Zhou et~al.(2024{\natexlab{b}})Zhou, Huang, Lu, Qin, and Wang}]{zhou2024don}
Yukai Zhou, Zhijie Huang, Feiyang Lu, Zhan Qin, and Wenjie Wang. 2024{\natexlab{b}}.
\newblock Don't say no: Jailbreaking llm by suppressing refusal.
\newblock \emph{arXiv preprint arXiv:2404.16369}.

\bibitem[{Zhu et~al.(2020)Zhu, Rawat, Zaheer, Bhojanapalli, Li, Yu, and Kumar}]{zhu2020modifying}
Chen Zhu, Ankit~Singh Rawat, Manzil Zaheer, Srinadh Bhojanapalli, Daliang Li, Felix Yu, and Sanjiv Kumar. 2020.
\newblock Modifying memories in transformer models.
\newblock \emph{arXiv preprint arXiv:2012.00363}.

\bibitem[{Zou et~al.(2023)Zou, Wang, Carlini, Nasr, Kolter, and Fredrikson}]{zou2023universal}
Andy Zou, Zifan Wang, Nicholas Carlini, Milad Nasr, J~Zico Kolter, and Matt Fredrikson. 2023.
\newblock Universal and transferable adversarial attacks on aligned language models.
\newblock \emph{arXiv preprint arXiv:2307.15043}.

\end{thebibliography}
